\documentclass[preprint,prd,showpacs,floatfix]{revtex4}

\usepackage{amsmath}
\usepackage{amsfonts}
\usepackage{amssymb}
\usepackage{bbm}
\usepackage{latexsym}
\usepackage[dvips]{graphicx}
\usepackage{graphics}
\usepackage{color}

\begin{document}

\title{Modified Friedmann Cosmologies -- theory \& observations} 
\author{Marek Szyd{\l}owski}
\email{uoszydlo@cyf-kr.edu.pl}
\affiliation{Astronomical Observatory, Jagiellonian University, Krak\'ow, Poland}
\author{Wojciech Czaja}
\email{czaja@oa.uj.edu.pl}
\affiliation{Astronomical Observatory, Jagiellonian University, Krak\'ow, Poland}
\date{\today}

\begin{abstract}
We investigate a class of Cardassian scenarios of the universe evolution in notions of the qualitative 
theory of dynamical systems. This theory allows us to analyze all solutions for all possible initial 
conditions on the phase plane. In the Cardassian models we find that big-rip singularities are present as
a typical behavior in the future if $n<0$. Some exact solutions for the flat Cardassian models as well 
as a duality relation were found. In turn from the statistical analysis of Knop's SNIa data, without 
any priors on matter content in the model, we obtain that at the $99\%$ confidence level this big-rip
scenario will reach. The potential function for the Hamiltonian description of dynamics is reconstructed 
from the SNIa data (inverse dynamical problem). We also pointed out the statistical analysis results 
depend oversensitively on the choice of the model parameter $\mathcal{M}$. 

\end{abstract}


\maketitle

\section{Introduction}

In this note we apply the qualitative analysis of differential equations to the Cardassian models which 
have become popular during last two years due to some of their interesting features. This class of models 
was proposed as an alternative to the cosmological constant model \cite{freese02,wang03} to explain the 
current acceleration of the Universe \cite{perlmutter99,riess98}. Freese and Lewis \cite{freese02} claimed 
that the Cardassian model, in which the Friedmann-Robertson-Walker (FRW) equation is modified, explains the 
expansion of the universe without any dark energy component. A certain additional term in the FRW 
equation which may arise from the fundamental physics (brane epoch) drives the present acceleration 
of the Universe. 

The Cardassian model survives several observational tests like the magnitude-redshift 
for the present type Ia supernovae data \cite{zhu03,zhu03b,avelino03,wang03,szydlowski03c,godlowski03}, 
$\theta$--$z$ test of the angular size of high-$z$ compact radio sources \cite{zhu02} or $d_{A}$--$z$ test
for the SZ/X-ray clusters proposed by Zhu and Fujimoto \cite{zhu03a}.

The main aim of this paper is to demonstrate the theoretical power in explanation of the cosmological 
problems by the Cardassian model. Moreover, the dynamical system methods allow to reveal some unexpected 
properties of this model. We construct the phase spaces for these models and discuss how their structure 
differs from the canonical model with the cosmological constant (Do new types of solutions appear? 
What is their physical meaning? Does any change of parameter values lead to some change of dynamical 
behavior?).

From the theoretical point of view it is interesting to analyze all evolutional paths of the Cardassian 
models for all initial conditions. On the other (empirical) hand it is important to check which 
Cardassian evolutional paths fit to astronomical observations, namely type Ia supernovae data. The 
both issues will be examined in this paper.

There is a widespread opinion that physically realistic models of the universe should possess some kind
of structural stability---the existence of too many dramatically different mathematical models which 
agree with observations would be fatal for the empirical method of science \cite{thom77}. A dynamical system 
is said to be structurally stable if other dynamical systems which are close to it (in a metric sense) are 
topologically equivalent (i.e. modulo homeomorphism). Although the question how to ensure such stability is
an open problem for the dynamical systems of higher dimension than two \cite{smale80} in the case of 
two-dimensional dynamical systems following the Peixoto theorem that states the structurally stable systems
form open and dense subsets in the space of all dynamical systems on the plane \cite{peixoto62}. Moreover, in the case of 
two-dimensional systems there is a simple test of structural stability. Namely, if the right-hand sides of 
the dynamical systems are in the polynomial form, the global phase portraits are structurally stable if the 
number of critical points and limit cycles is finite, each point is hyperbolic and there are no trajectories
connecting saddle points. 

In the case considered the dynamics is reduced to the form of a two dimensional dynamical system with 
right-hand sides in the polynomial form
\begin{align}
\dot{x} \equiv \frac{dx}{dt} = P(x,y),\\
\dot{y} \equiv \frac{dy}{dt} = Q(x,y),
\label{eq:1}
\end{align}
where $P, Q \in C^{\infty}$ class of functions; $(x,y)$ is a differential space $\mathcal{M}$, which is
useful for visualization of the dynamics, called the phase space (or state space). The right-hand 
sides define a vector field $\mathcal{V}=(P,Q)$ belonging to the tangent bundle $\mathcal{M}$.
Now we can define a phase curve as a integral curve of the vector field. All phase curve with the critical 
points $(P(x_{0},y_{0})=0$, $Q(x_{0},y_{0})=0)$ constitute the phase portrait of the system. 
Two phase portraits are equivalent if there exists an orientation preserving a homeomorphism which maps 
integral curves of both systems into each other. 

From the physical point of view a critical point represents asymptotic states or equilibria. The main aim of 
qualitative analysis of differential equations is constructing the phase portrait of the system. 
Following the Hartman-Grobman theorem: the nonlinear dynamics near the hyperbolic critical points
($\forall i$ ${\rm Re} \lambda_{i} \neq 0$, where $\lambda_{i}$ is an eigenvalue of a linearization matrix)
is qualitatively equivalent to its linear part
\begin{align}
\label{eq:2}
\dot{x}&=\frac{\partial P}{\partial x}(x_{0},y_{0})(x-x_{0})
+\frac{\partial P}{\partial y}(x_{0},y_{0})(y-y_{0}) \\
\label{eq:3}
\dot{y}&=\frac{\partial Q}{\partial x}(x_{0},y_{0})(x-x_{0})
+\frac{\partial Q}{\partial y}(x_{0},y_{0})(y-y_{0}).
\end{align}

Full knowledge of the dynamical system comprises also its behavior at infinity. To achieve this one
usually transforms the phase plane into a Poincar\'e sphere. Then infinitely distant points of the plane 
are mapped into the equator of the sphere. Of course, the character of critical points (which depends on 
the solutions of characteristic equation $\lambda^{2}-\lambda{\rm Tr}\mathcal{A}+\det\mathcal{A}=0$ for 
linearization matrix $\mathcal{A}$) is conserved but new critical points can appear at the equator. Now, 
the orthogonal projection of any hemisphere onto the tangent plane gives the compactified portrait on 
compact projective plane.

There are two main aims of the paper. First, theoretical investigations of the dynamics and second, 
its reconstruction from Knop's SNIa sample (inverse dynamical problem). We found that the Cardassian 
model very well fits the SNIa data and that its unexpected future (the big-rip singularity) is consistent 
with SNIa data on the $99\%$ confidence level without any priors on matter content. 
We have demonstrated how some information about dark 
energy can be deduced from the potential of the Hamiltonian system describing the evolution of 
the universe. Using the potential function instead of the equation of state parameter $w=p/\rho$ 
seems to be attractive in the context of searching the adequate description of the present stage 
of evolution of the Universe.

\section{Qualitative cosmology of the Cardassian model}

\subsection{Dynamics of Cardassian models filled with single fluid on the $(H,\rho)$ phase plane}

Let us consider the Friedmann equation generalized by Freese and Lewis \cite{freese02} to the form
\begin{equation}
H^{2}=\frac{\rho}{3}+f(\rho)-\frac{k}{a^{2}}.
\label{eq:4}
\end{equation}

Let us note that equation (\ref{eq:4}) is the first integral of the generalized Einstein equations
for the Robertson-Walker metric. The first of these equations is called the Raychaudhuri equation 
and second one is called the conservation equation
\begin{align}
\dot{H}&=-\frac{1}{2}\big(1+3f'(\rho)\big)(\rho+p)+\frac{1}{3}(\rho+3f(\rho))-H^{2},\nonumber \\
\dot{\rho}&=-3H(\rho+p),
\label{eq:5}
\end{align}
where we assume that the matter content is described by the hydrodynamical energy-momentum tensor with
energy density $\rho$ and pressure $p$.

\begin{figure}[!ht]
\begin{center}
$\begin{array}{c@{\hspace{0.2in}}c@{\hspace{0.2in}}c}
\multicolumn{1}{l}{\mbox{\bf (a)}} & 
\multicolumn{1}{l}{\mbox{\bf (c)}} &
\multicolumn{1}{l}{\mbox{\bf (e)}} \\ [0.4cm]
\includegraphics[scale=0.35]{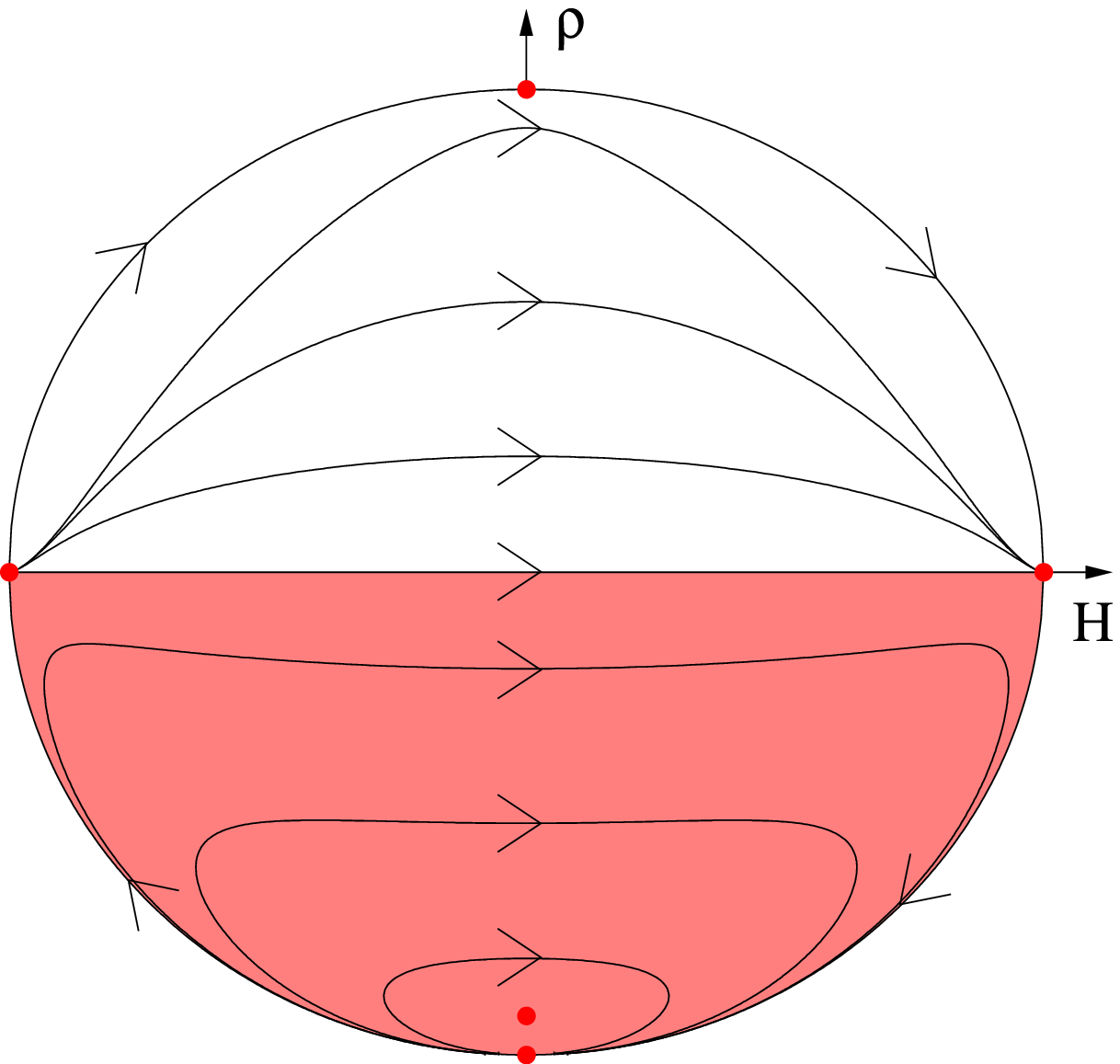}  & 
\includegraphics[scale=0.35]{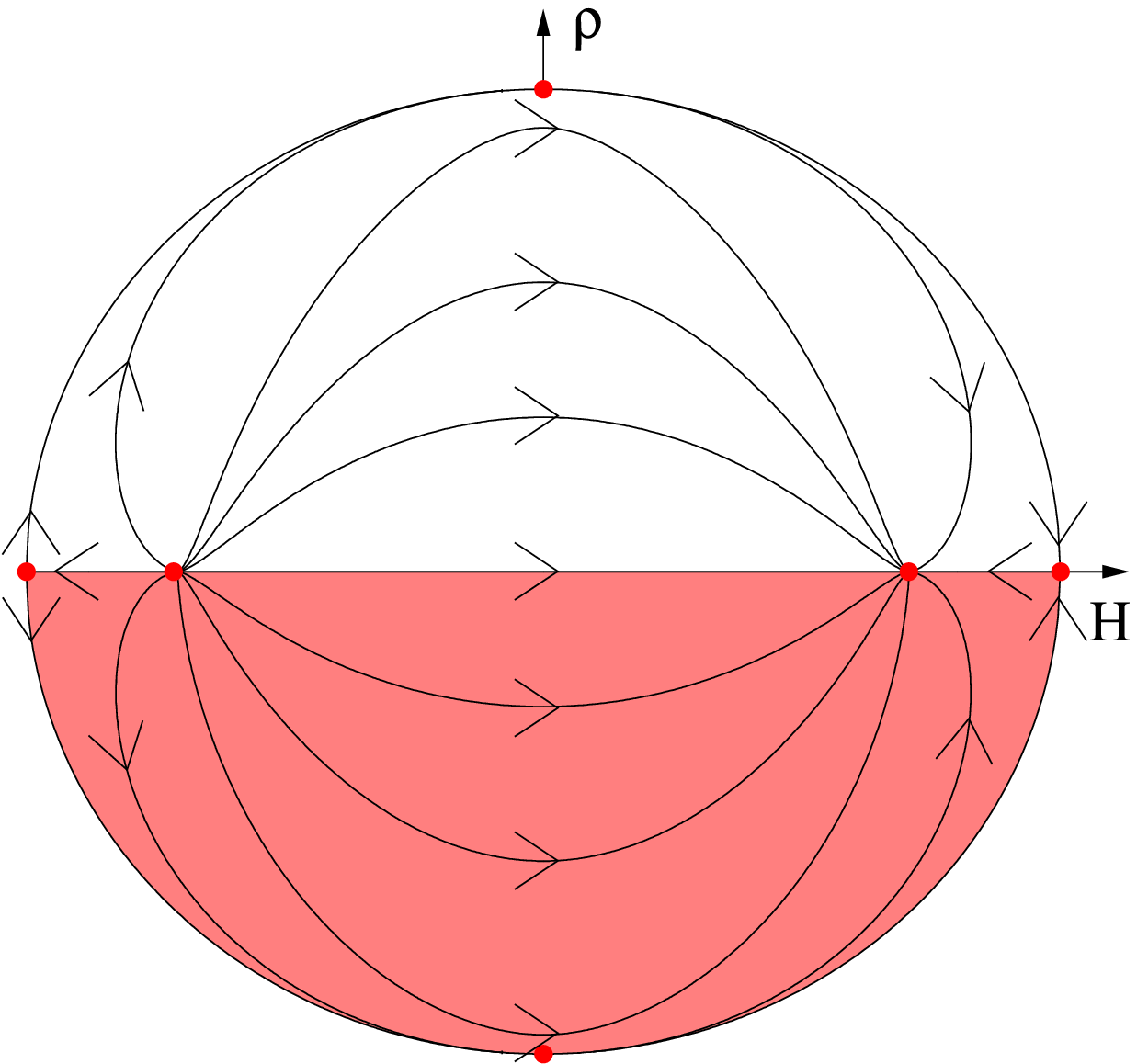} &
\includegraphics[scale=0.35]{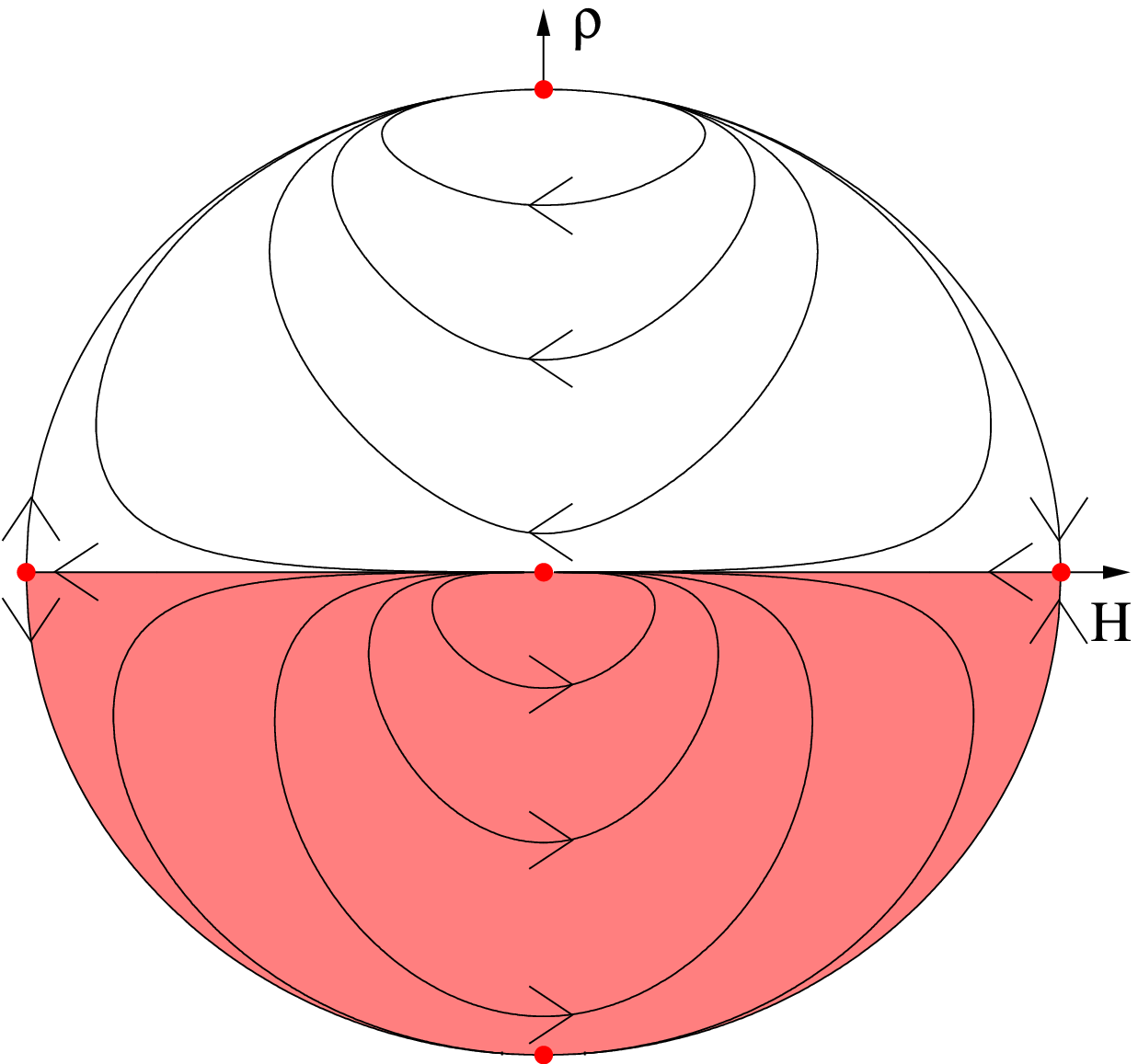} \\ [0.0cm]
\multicolumn{1}{l}{\mbox{\bf (b)}} & 
\multicolumn{1}{l}{\mbox{\bf (d)}} &
\multicolumn{1}{l}{\mbox{\bf (f)}} \\ [0.4cm]
\includegraphics[scale=0.35]{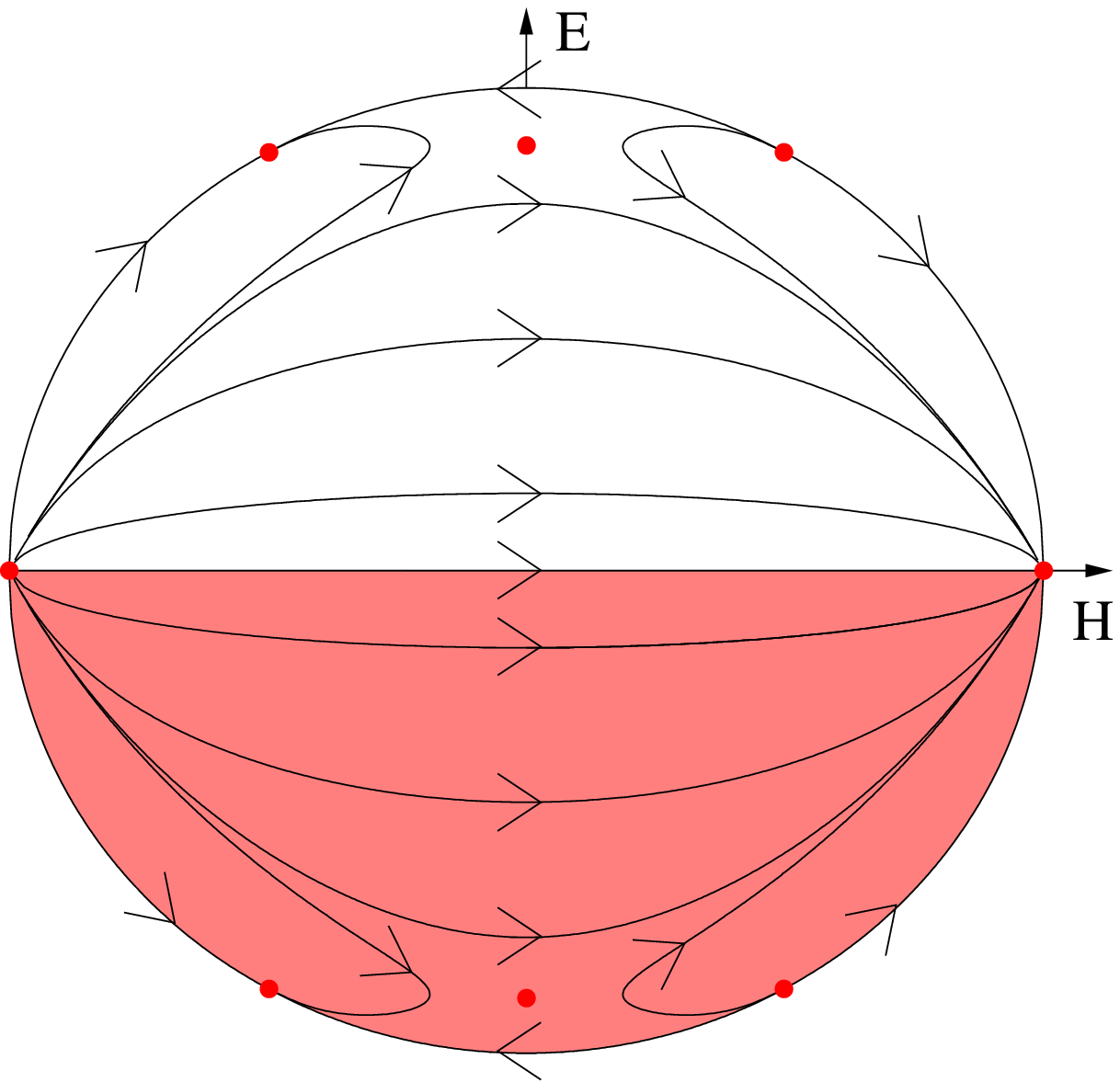}  & 
\includegraphics[scale=0.35]{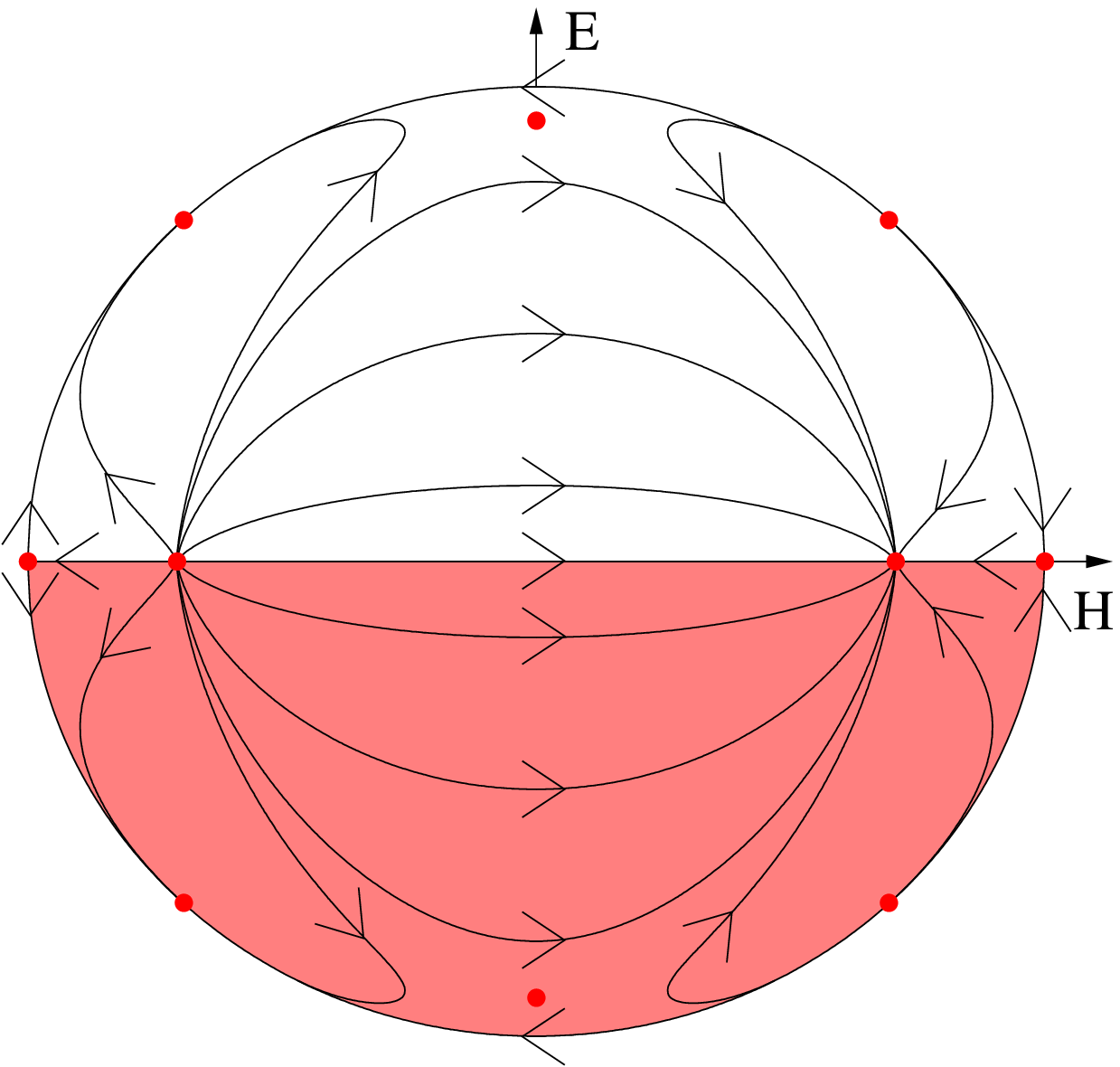} &
\includegraphics[scale=0.35]{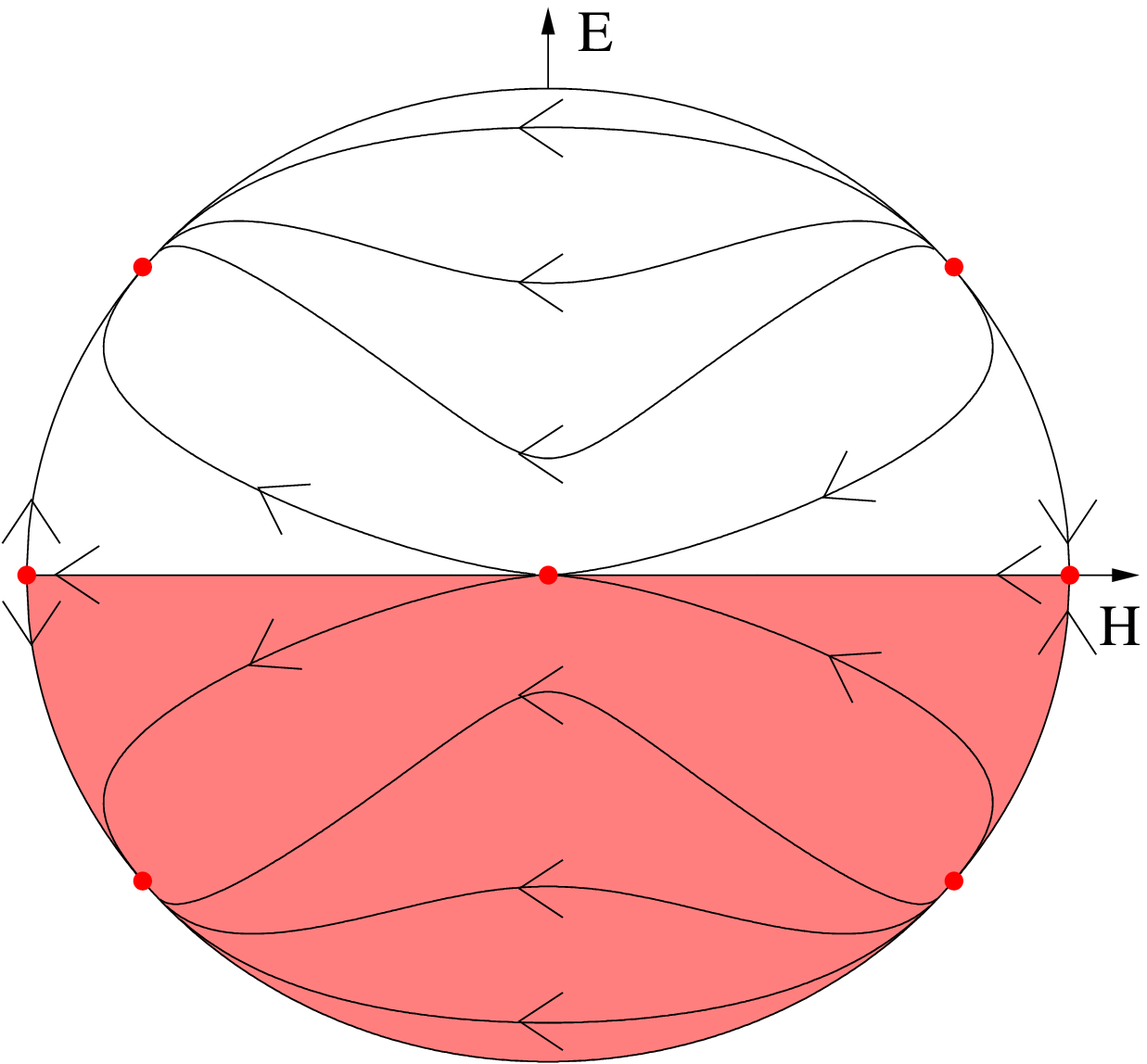} \\ [0.0cm]
\end{array}$
\end{center}
\caption{The phase portraits for Cardassian scenarios described by the systems of differential equations 
(\ref{eq:6}),(\ref{eq:7}) with different values of parameter $n$: {\bf (a)}, {\bf (b)} $n=-1$; {\bf (c)},
{\bf (d)} $n=0$; {\bf (e)}, {\bf (f)} $n=1$. Two types of critical points can exist at a finite domain of 
the phase space: 1) static $H_{0}=0$ and $E=E_{0}$ only if $n<2/3$ or 2) non-static if 
$H_{0} \neq 0$, $E_{0}=0$, hence $n=0$ and $H_{0}=\pm \sqrt{B}$ (cosmological constant case). If $n>0$ only 
($0,0$) critical point is admissible. The static critical points represent the static universe in Cardassian
models while stable and unstable nodes for $n=0$ are de Sitter solutions. The points at infinity can 
principally be of two types. First, if asymptotically standard contribution from the $\rho$ dominates near 
the initial singularity, and the second one if the Cardassian term dominates the future evolution (future 
singularity of type $a \rightarrow \infty$, $H \rightarrow \infty$, $E \rightarrow 0$). All these points lie
on the intersection of the trajectory of the flat model with a circle at infinity. Future singularities, 
called big-rip singularities, in the original time variable, correspond to a finite value of $t=t_{0}$. 
We can find it on the phase plane as an intersection of a circle at infinity with line 
$\{(H,E)\colon H=\pm E^{n}\sqrt{B(2-3n)/2}, E \rightarrow 0$\}. Note that models with $n>0$ are structurally
unstable because of degenerated critical points at a finite domain of phase space.
}
\label{fig1}
\end{figure}

For the universe filled with dust ($p=0$) system (\ref{eq:5}) takes the form
\begin{gather}
\dot{H} = -\frac{1}{6}\rho+\bigg(1-\frac{3}{2}n\bigg)B\rho^{n}-H^{2},\nonumber \\
\dot{\rho} = -3H\rho,
\label{eq:6}
\end{gather}
where $f(\rho)=B\rho^{n}$.\\
After the transformation $\rho \rightarrow E = \sqrt{\rho}$ we obtain an equivalent dynamical system
\begin{gather}
\dot{H} = -\frac{1}{6}E^{2}+\bigg(1-\frac{3}{2}n\bigg)BE^{2n}-H^{2},\nonumber \\
\dot{E} = -\frac{3}{2}HE.
\label{eq:7}
\end{gather}
The phase portraits for Cardassian scenarios described by systems (\ref{eq:6}) and (\ref{eq:7})
(for different values of $n$) are presented in Fig.~\ref{fig1}.

System (\ref{eq:7}) has the first integral in the form
\begin{equation}
\rho_{\rm eff}=\rho+3B\rho^{n}=E^{2}+3BE^{2n}=3H^{2}+\frac{3k}{a^{2}}.
\label{eq:7b}
\end{equation}
Note that this system is invariant with respect to reflection symmetry $E \rightarrow -E$. 
As a consequence of it we obtain symmetric phase portraits.

Let us now consider the critical points of system (\ref{eq:7}) at a finite domain of phase space 
$\{(H,E)\colon E \geq 0\}$. They can be of two types: 1) static ($H_{0}=0$, $E_{0} \neq 0$) 
or 2) non-static ($H_{0} \neq 0$, $E_{0} = 0$). The static critical point exists if $n \leq 2/3$ only.
Then
\begin{equation}
E_{0}=[3B(2-3n)]^{\frac{1}{2(1-n)}},\quad H_{0}=0.
\label{eq:7c}
\end{equation}
In the opposite case if $n>2/3$ there are no static critical points at a finite domain of phase space.
The non-static critical points must be located on the line $\{E=0\}$ ($H$-axis). Therefore in this case
$n \geq 0$ is required ($n>1/2$ for a vector field to be class $C^{1}$). And we can find that it is possible
only for $H_{0}=0$.

From the physical point of view the static critical points represent static universes while 
non-static critical points which are admissible only for $n \geq 0$ represent an empty Minkowski solution.
For discussion of the character of the critical points it is convenient to use the linearization matrix at 
the critical points and the characteristic equation. In the case considered we have
\begin{equation}
\mathcal{A} = \left[
\begin{matrix}
-2H_{0},& -\frac{1}{3}E_{0}+n(2-3n)BE_{0}^{2n-1}\\
-\frac{3}{2}E_{0},& -\frac{3}{2}H_{0}\\
\end{matrix}
\right]. 
\label{eq:7d}
\end{equation}
Hence for non-static critical points ($n>1/2$) we always obtain degenerated critical point while for static
critical points the characteristic equation takes the form
\begin{equation}
\lambda^{2}-\lambda({\rm Tr}\mathcal{A})+{\rm \det{\mathcal{A}}}=0,
\label{eq:7e}
\end{equation}
where ${\rm Tr}\mathcal{A}=0$ and $\det{\mathcal{A}}=1/2E_{0}(n-1)$.
Therefore, if $n<1$, then the critical points are of a saddle type because the eigenvalues of the 
linearization matrix are real and opposite signs. In the special case $n=1$, the static critical point 
does not exist (see Fig.~\ref{fig1}) and non-static critical point is degenerated similarly to the case 
$n>1/2$. This means that they are corresponding systems because of the presence of non-hyperbolic and 
degenerated critical points in the phase space.

The behavior of trajectories near the non-degenerated static critical points is equivalent to the behavior 
of linearized system. The linearized system (\ref{eq:7}) in the vicinity of this point is
\begin{align}
\dot{H}&=\frac{1}{3}E_{0}(n-1)[E-E_{0}], \nonumber \\
\dot{E}&=-\frac{3}{2}E_{0}H,
\label{eq:7f}
\end{align}
where $E_{0}$ is given by (\ref{eq:7c}) and this point is structurally stable (saddle point) because
the eigenvalues of the linearization matrix have opposite signs.

After integration of the system (\ref{eq:7f}) we obtain
\begin{align}
H&=c_{1}e^{-\lambda_{1}t}+c_{2}e^{\lambda_{2}t}, \nonumber \\
E-E_{0}&=c_{1}k_{1}e^{\lambda_{1}t}+c_{2}k_{2}e^{\lambda_{2}t},
\label{eq:7g}
\end{align}
where $\lambda_{1,2} = \pm \sqrt{\det{\mathcal{A}}} = \pm E_{0} \sqrt{(1-n)/2}$ and $k_{1}$, $k_{2}$
are the components of the eigenvectors determining main directions along which particular trajectories
approach or escape from the critical point.

For completeness of our analysis of the dynamics it is important to investigate the behavior of trajectories
at infinity. For this aim it is useful to compactify $\mathbb{R}^{2}$ phase space by adding a circle at 
infinity. One can simply do that by introducing of the projective coordinates. Two projective maps $(z,u)$, 
$(v,w)$ cover a circle at infinity given by 
$$ z=0,\quad -\infty<v<\infty \quad (H=\infty,\quad z=1/H,\quad u=E/H)$$
and
$$ v=0,\quad -\infty<w<\infty \quad (E=\infty,\quad v=1/E,\quad w=H/E).$$
The system under consideration in the projective coordinates has the form
\begin{align}
\frac{dz}{d\tau}&=z+\frac{1}{6}zu^{2}+\frac{1}{2}B(3n-2)u^{2n}z^{(3-2n)},\nonumber \\
\frac{du}{d\tau}&=-\frac{3}{2}u+u\bigg[1+\frac{1}{6}u^{2}+\frac{1}{2}B(3n-2)u^{2n}z^{2(1-n)}\bigg],
\label{eq:7h}
\end{align}
where $\tau$ is a new time variable such that $t \rightarrow \tau \colon d\tau = dt/z$ -- it is monotonic
function of original time variable $t$ only in the domain $z>0$. We find that ($z_{0}=0,u_{0}^{2}=3$)
($n<2/3$) is a critical point of the system (\ref{eq:7h}) and for $n \geq 1/2$ we obtain additionally
the critical point ($z_{0}=0,u_{0}=0$). In this case we have
\begin{equation*}
{\rm Tr}\mathcal{A}=\frac{1}{2}+\frac{2}{3}u_{0}^{2},\quad \det{\mathcal{A}}=\frac{1}{12}\Big(6+u_{0}^{2}\Big)\Big(u_{0}^{2}-1\Big).
\end{equation*}
After simple calculation we obtain that the critical points ($0,\pm \sqrt{3}$) are stable nodes located
on a circle at infinity as intersection points with line $E_{0}=\pm H_{0}$. The eigenvalues of the 
linearization matrix at the point ($z_{0},u_{0}=\pm\sqrt{3}$) are $\lambda_{1}=1$ and $\lambda_{2}=3/2$
i.e., they are real and positive therefore the critical points are unstable (${\rm Tr}\mathcal{A}=5/2>0$)
for both $H_{0}>0$ and $H_{0}<0$ if considered in the original time parameterization (see Fig.~\ref{fig1}).
The dynamical system methods offer the possibility of a simple detection of some other unexpected 
critical points in $\Lambda$CDM model. Note that the critical points at infinity can also appear on 
the intersection of a circle at infinity with lines $\{(H,E)\colon H=\pm E^{n}\sqrt{B(2-3n)/2}\}$. 
Then $E \rightarrow 0$ and $H \rightarrow \infty$ only if $n$ assumes negative values. This critical point 
represents an asymptotic state of the evolution dominated by the Cardassian term as $E \rightarrow 0$ 
($a \rightarrow \infty$). As a result of such domination for $n<2/3$ we obtain dangerous big-rip
singularities. From Fig.~1. one can conclude that if $n<0$ they are typical asymptotic stages of the
evolution in the future. In this stage $H$ is going to infinity while energy $E$ is reaching zero.
This critical point corresponds to $\tau = \ln{a} = \infty$ or a finite value of the cosmological time $t$.
In the projective coordinates at this point we have
\begin{equation}
\frac{z^{n-1}}{u^{n}} = \pm \sqrt{B\frac{2-3n}{2}},\quad u=0.
\end{equation}

\begin{figure}[!ht]
\begin{center}
$\begin{array}{c@{\hspace{0.2in}}c}
\multicolumn{1}{l}{\mbox{\bf (a)}} & 
\multicolumn{1}{l}{\mbox{\bf (b)}} \\ [-0.5cm]
\includegraphics[scale=0.32, angle=270]{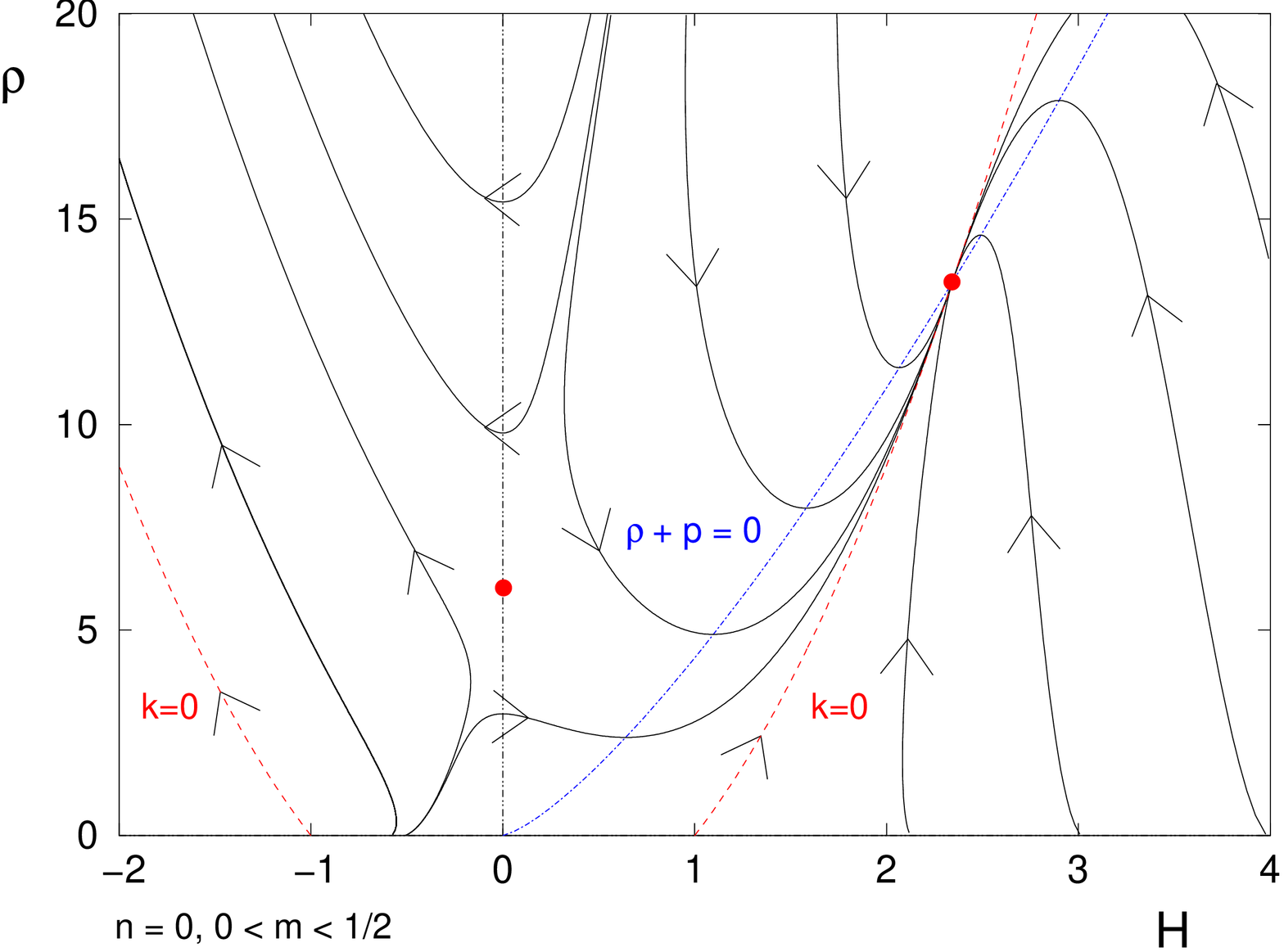} & 
\includegraphics[scale=0.32, angle=270]{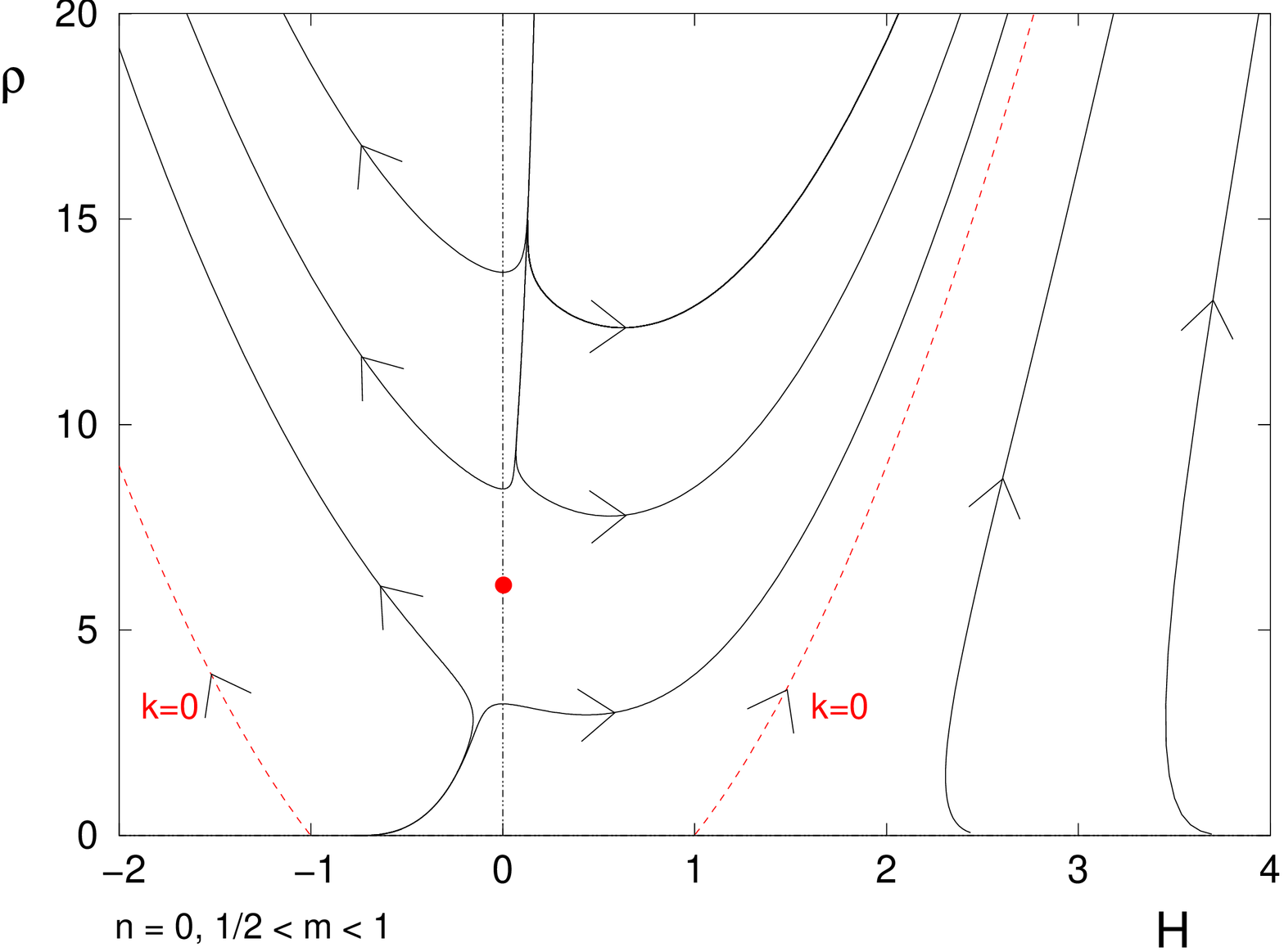} \\ [0.4cm]
\multicolumn{1}{l}{\mbox{\bf (c)}} & 
\multicolumn{1}{l}{\mbox{\bf (d)}} \\ [-0.5cm]
\includegraphics[scale=0.32, angle=270]{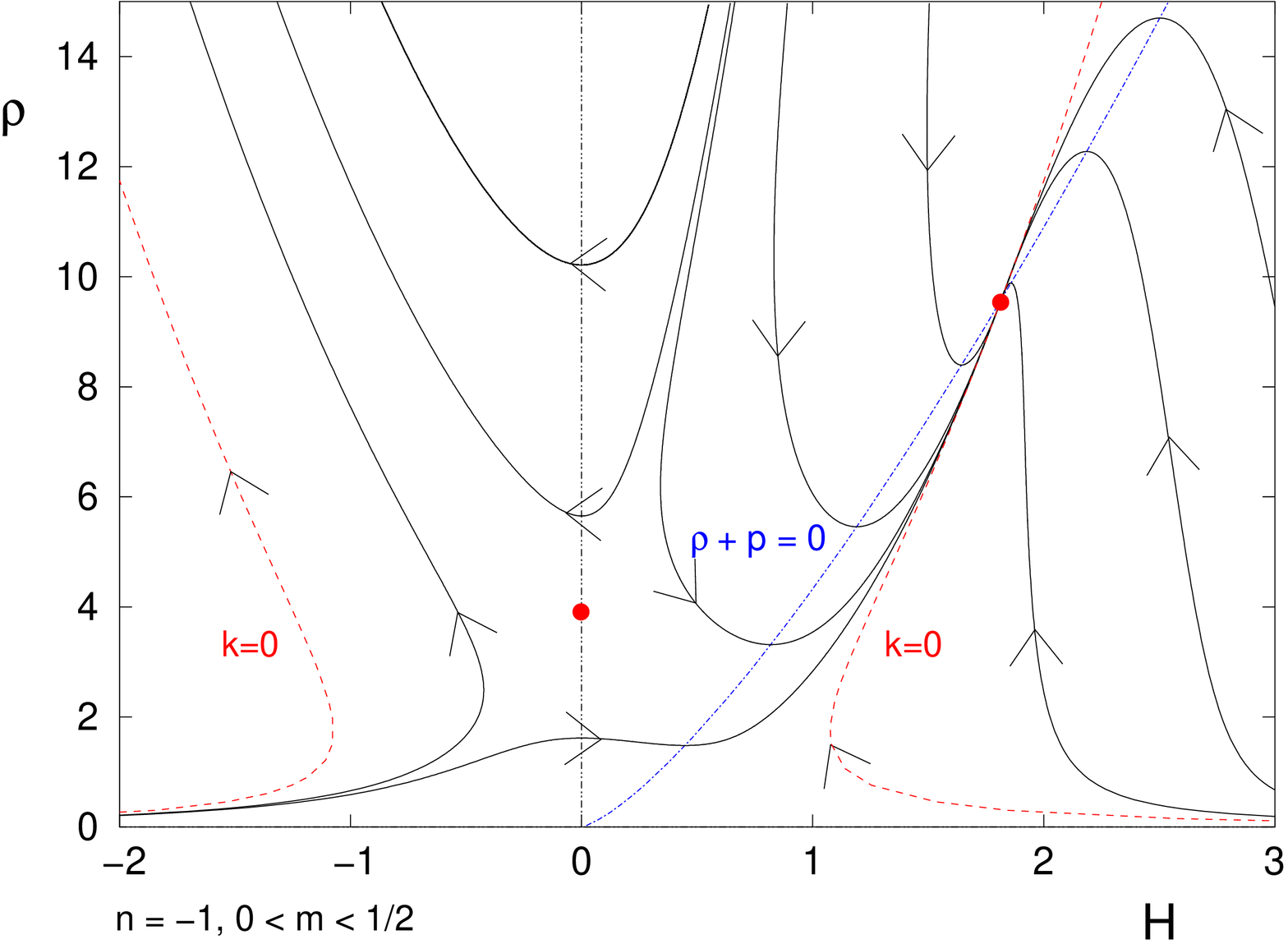} & 
\includegraphics[scale=0.32, angle=270]{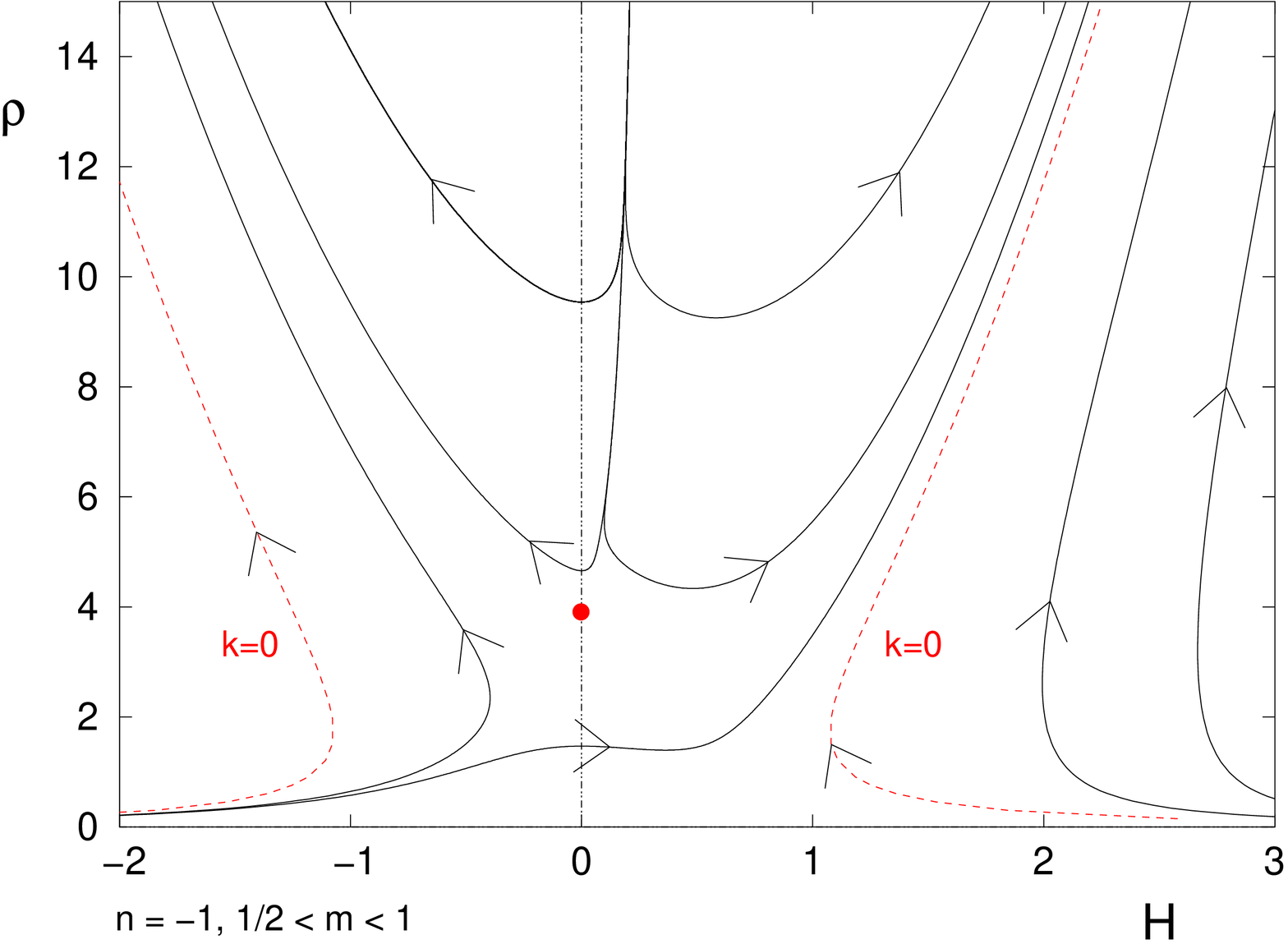} \\ [0.4cm]
\end{array}$
\end{center}
\caption{The phase portraits for Cardassian scenarios described by the system of differential equations 
(\ref{eq:7a}) for $B=\alpha=1$ and for different values of parameters ($n,m$). There are critical points of
two types: static which lie on the $\rho$-axis and non-static which are the intersection of the trajectory 
of the flat model (dashed line) with the boundary of the weak energy condition $\rho=3\alpha \rho^{m}H$ 
(dotted line). Note that obtained phase portraits are structurally stable at a finite domain because of the 
presence of nodes and saddles.}
\label{fig1a}
\end{figure}

The phase plane $(H,\rho)$ can also be useful in analysis of model under consideration in which the 
effects of bulk viscosity are included. They are equivalent to introducing the effective pressure 
$p_{\rm eff}=p-3 \xi H$ with exact dependence on the Hubble function; here $\xi$ is called viscosity coefficient
and in the Belinskii parameterization takes the form $\xi = \alpha \rho^{m}$. 
For this case we can rewrite system~(\ref{eq:5}) in the form
\begin{gather}
\dot{H} = -\frac{1}{2}\bigg(\frac{1}{3}\rho-3 \alpha \rho^{m}H\bigg)+\bigg[1-\frac{3}{2}n\big(1-3\alpha \rho^{m-1}H\big)\bigg]B\rho^{n}-H^{2},\nonumber \\
\dot{\rho} = -3H(\rho-3\alpha \rho^{m}H).
\label{eq:7a}
\end{gather}
The phase portraits for representative cases are demonstrated in Fig.~\ref{fig1a}.

As it is well known there are some major motivations to study such dark energy models which have de Sitter 
solution as a global attractor at late times \cite{hao03}. From Fig.~\ref{fig1a} we can observe the 
existence of de Sitter solution as a global attractor. Note that there are both open 
and closed universes approaching this critical point. All such critical points are the solutions of the 
following system
\begin{gather}
\rho_{\rm eff} -  3H^{2} = 0 \quad {\rm or} \quad \rho + 3B \rho^{n} = 3H^{2},\nonumber \\
\rho =3 \xi(\rho) H = 3 \alpha \rho^{m} H.
\label{eq:7i}
\end{gather}
The analysis of the evolution of FRW cosmological models with dissipation, by using ($H,\rho$) plane,
can be performed for the general form of quintessence coefficient $w = w(\rho, H)$, $p = w \rho$, or for the 
models with causal viscosity \cite{prisco01}.

While the evolution of Cardassian model of the universe filled with single fluid, for which the equation 
of state assumes the form $p=w_{i}\rho$, is described by the system of differential equations in the phase 
variables $(H,\rho)$, it is useful to consider a different representation of the dynamics basing on
the Hamilton methods.

\subsection{Particle--like description of Cardassian models}

In our previous works \cite{szydlowski03,szydlowski03c} it was shown that the dynamics of any quintessential
FRW cosmologies can be reduced to the form of a one-dimensional Hamiltonian flow. In this approach we postulate 
that the equation of state has the general form $p=w(a(z))\rho$ in which the equation of state factor $w$ can 
be parameterized by the scale factor $a$ or redshift $z$. We obtain that any FRW dynamics with such a fluid
is described by the system of differential equations
\begin{align}
\dot{x}&=y,\nonumber \\
\dot{y}&=-\frac{\partial V(x)}{\partial x},
\label{eq:8}
\end{align}
where $x=a/a_{0}$, $a$ is the scale factor of the universe, and the Hamiltonian of the system under 
consideration assumes
\begin{equation}
\mathcal{H}=\frac{y^{2}}{2}+V(x)=0.
\label{eq:8a}
\end{equation}
The potential function $V(x)$ for system~(\ref{eq:8}) takes the form
\begin{equation}
V(x)=-\frac{\rho_{\rm eff}x^{2}}{6}.
\label{eq:9}
\end{equation}
For standard Cardassian model of the universe, filled with matter, for which the equation of state
is $p=w_{i}\rho$ ($w_{i}={\rm const.}$), we can simply obtain that
\begin{equation}
V(x)=-\frac{1}{2}\bigg\{\Omega_{i,0}x^{-1-3w_{i}}+\Omega_{{\rm Card},0}x^{-1-3\gamma}+\Omega_{k,0}\bigg\},
\label{eq:9a}
\end{equation}
where $\gamma=n(1+w_{i})-1$.

The advantage of representing dynamics in terms of Hamiltonian is the possibility to discuss the 
stability of critical points only from the geometry of the potential function. In the general case of the 
system with natural Lagrangian function, if a diagram of the potential function has a maximum and is upper 
convex, then it corresponds to the stable attractor. In our case the Hamiltonian function takes the 
special form of a simple mechanical system. Then the linearization matrix $\mathcal{A}$ is such that:
\begin{equation}
{\rm Tr}\mathcal{A}=0, \quad {\rm and} \quad \det{\mathcal{A}}=\Bigg(\frac{\partial^{2}V}{\partial x^{2}}\Bigg)\Bigg|_{(x=x_{0},y=0)}.
\label{eq:9b}
\end{equation}
Therefore from the characteristic equation $\lambda^{2}-\lambda({\rm Tr}\mathcal{A})+\det{\mathcal{A}}=0$
we obtain that eigenvalues of the linearization matrix should be 1) real and opposite signs or 2) purely
imaginary and conjugated. Therefore the possible critical points in a finite domain of the phase space
are only saddles (for the first case) or centers (for the second case). In the case considered we have
\begin{gather}
x_{0} = \Bigg[-\frac{1+3 \gamma}{1+3w_{i}} \frac{\Omega_{\rm Card}}{\Omega_{i,0}}\Bigg]^{\frac{1}{3(\gamma - w_{i})}},\nonumber \\
\det{\mathcal{A}} = -\frac{3}{2} x_{0}^{-3(1+w_{i})}\Omega_{i,0} (1+w_{i})(n+1)(1+3w_{i}).
\label{eq:9c}
\end{gather}
Therefore $w_{i} = -1,-1/3$; $n=-1$ are always bifurcations values of model parameters for which the global
structure of the phase space changes qualitatively. If critical point exists ($(1+ 3\gamma)(1+3w_{i}) < 0$)
then on the phase plane ($x, \dot{x}$) we obtain saddles at a finite domain if only 
$(1+w_{i})(1+3w_{i})(n+1) > 0$. In the opposite case we obtain centers which are structurally unstable.

\begin{figure}[ht]
\begin{center}
\includegraphics[scale=0.7, angle=0]{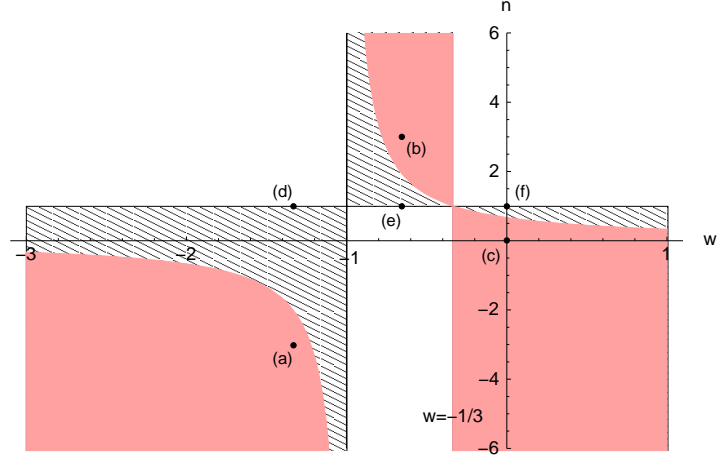}
\end{center}
\caption{Domains of parameters on the plane $(w,n)$ for which the Cardassian models described by the system
of differential equations (\ref{eq:8}) start to accelerate for $a>0$. In these cases the potential function 
$V(a)$ of the system has a maximum for $a>0$ and is upper convex.}
\label{fig2}
\end{figure}

\begin{figure}[ht]
\begin{center}
\includegraphics[scale=0.4, angle=270]{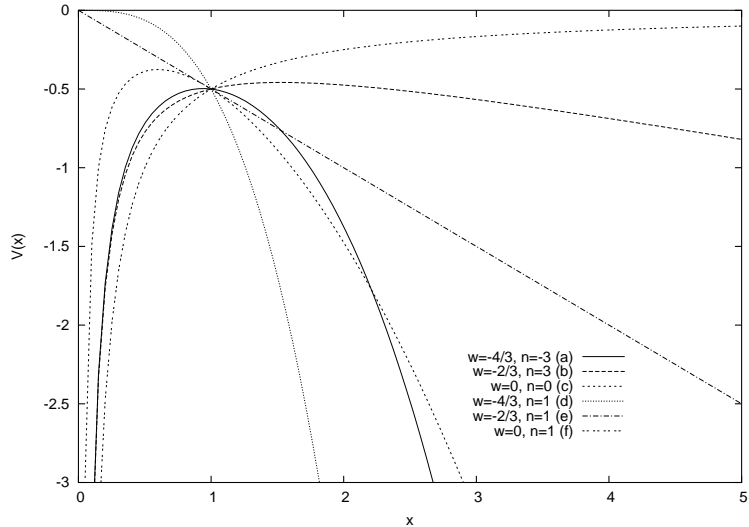} 
\end{center}
\caption{Plot of the potential functions describing Cardassian evolution of the universe for different 
parameters $(w,n)$ chosen and signed in Fig.~\ref{fig2} as {\bf (a)}-{\bf (f)}.}
\label{fig3}
\end{figure}

\begin{figure}[!ht]
\begin{center}
$\begin{array}{c@{\hspace{0.2in}}c@{\hspace{0.2in}}c}
\multicolumn{1}{l}{\mbox{\bf (a)}} & 
\multicolumn{1}{l}{\mbox{\bf (c)}} &
\multicolumn{1}{l}{\mbox{\bf (e)}} \\ [0.4cm]
\includegraphics[scale=0.35]{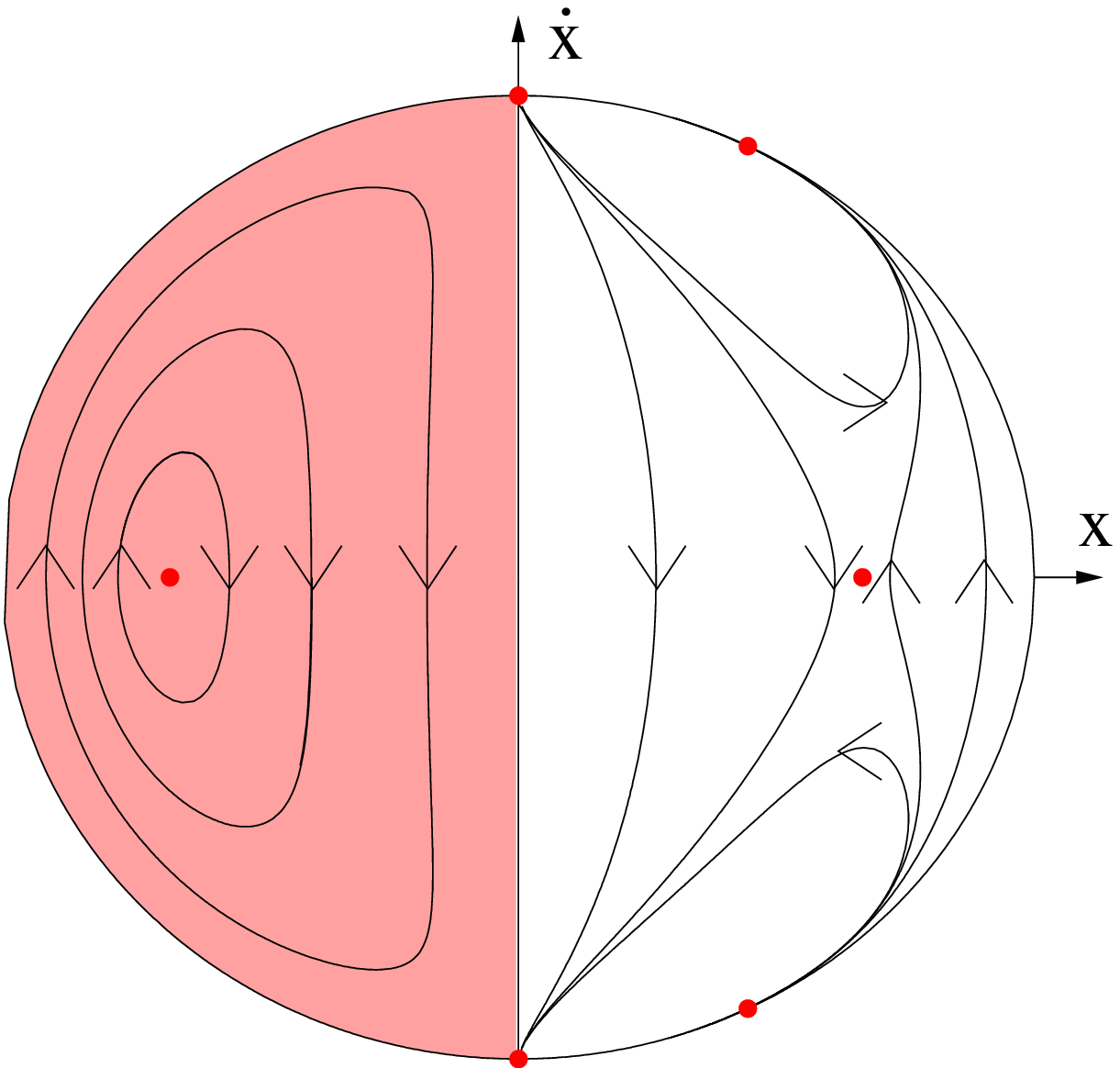}  & 
\includegraphics[scale=0.35]{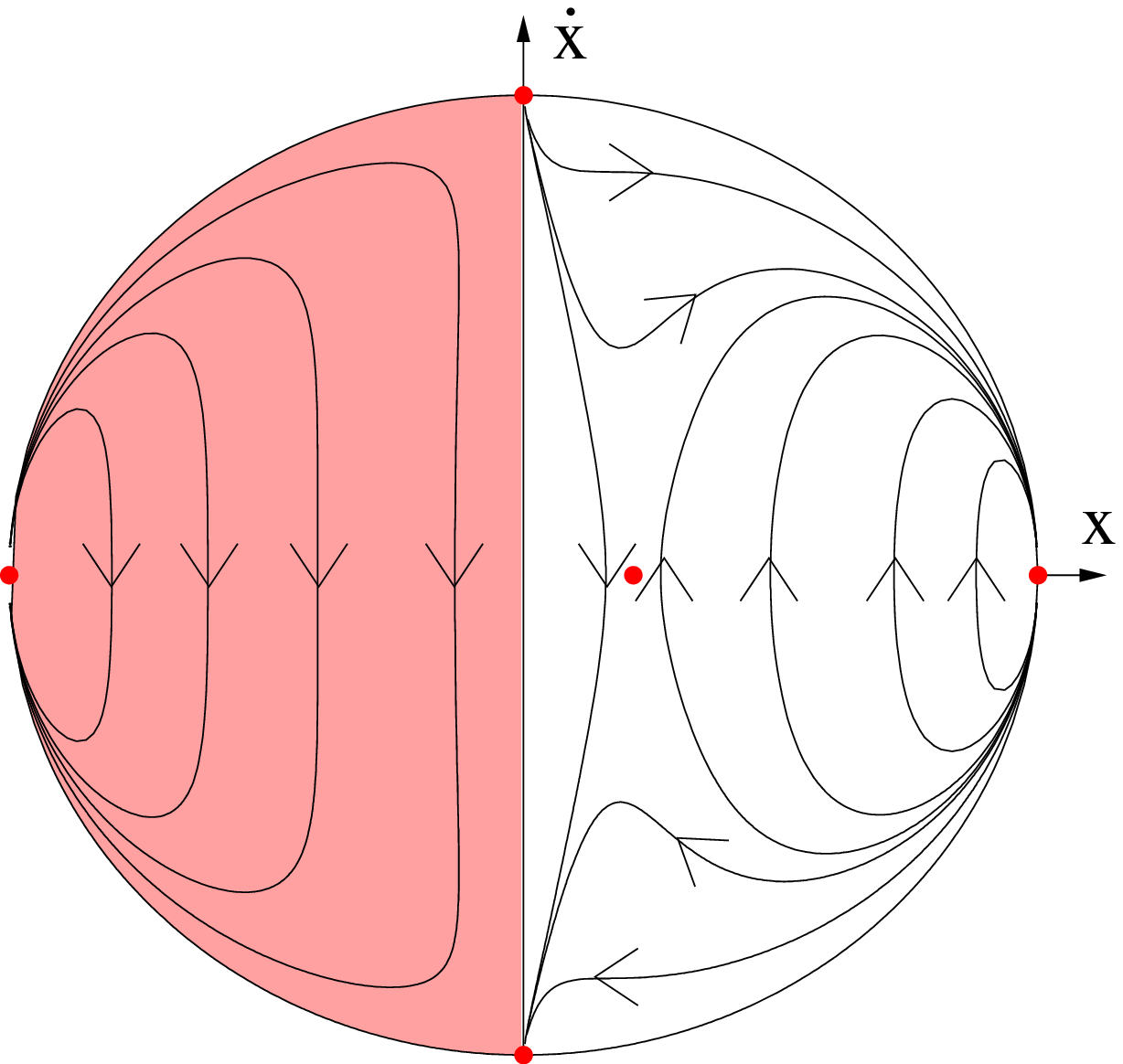} &
\includegraphics[scale=0.35]{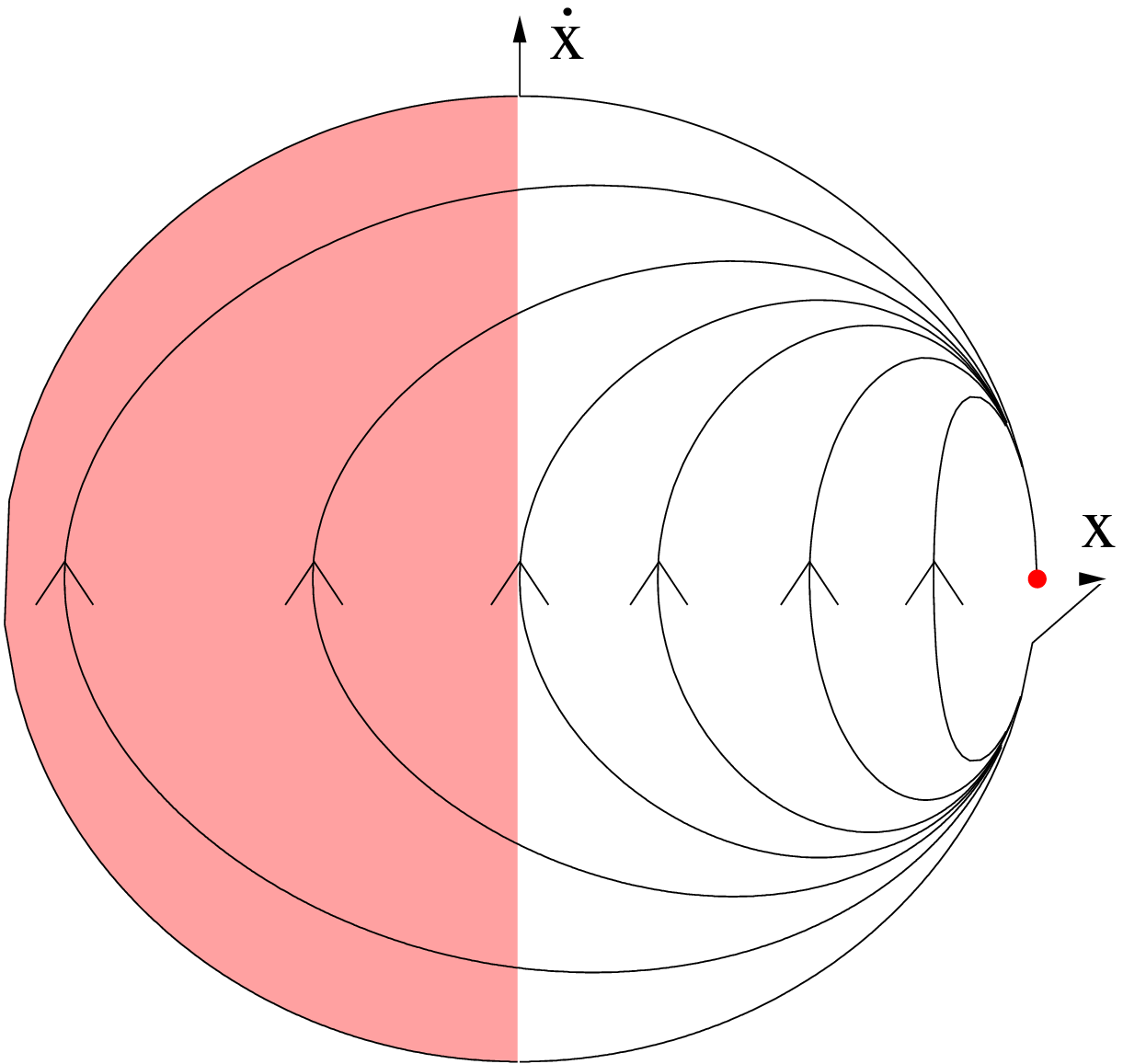} \\ [0.0cm]
\multicolumn{1}{l}{\mbox{\bf (b)}} & 
\multicolumn{1}{l}{\mbox{\bf (d)}} &
\multicolumn{1}{l}{\mbox{\bf (f)}} \\ [0.4cm]
\includegraphics[scale=0.35]{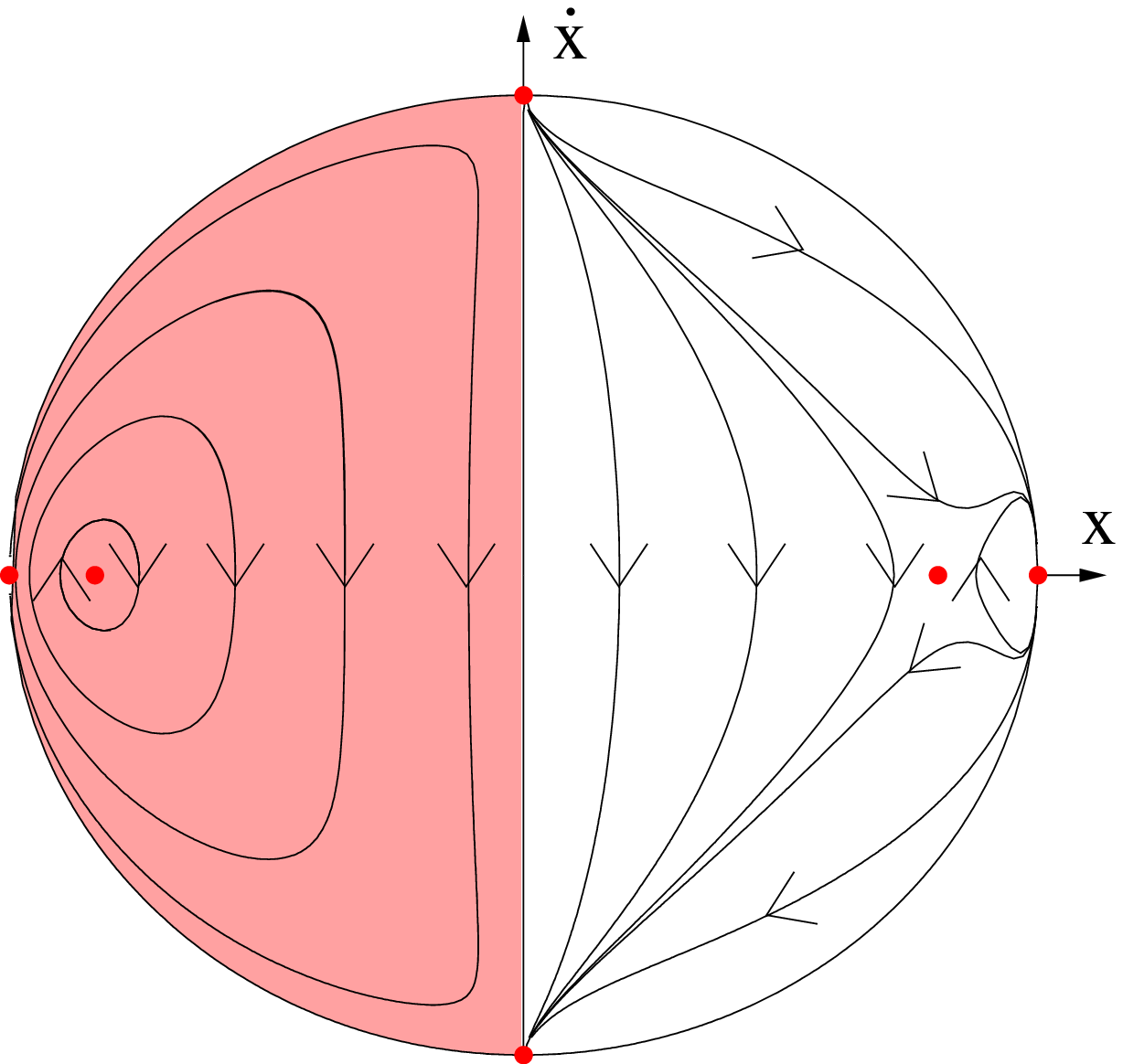}  & 
\includegraphics[scale=0.35]{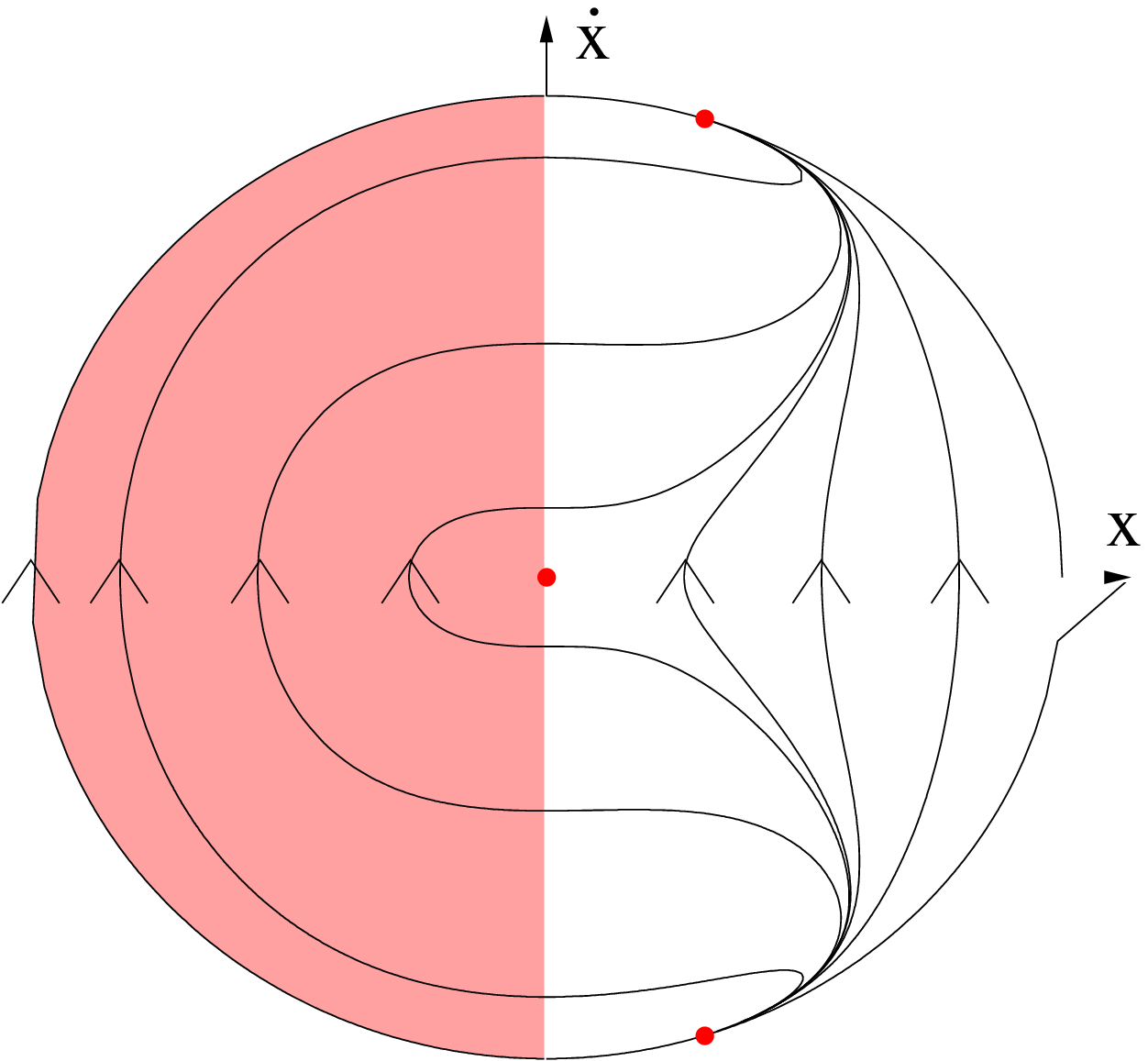} &
\includegraphics[scale=0.35]{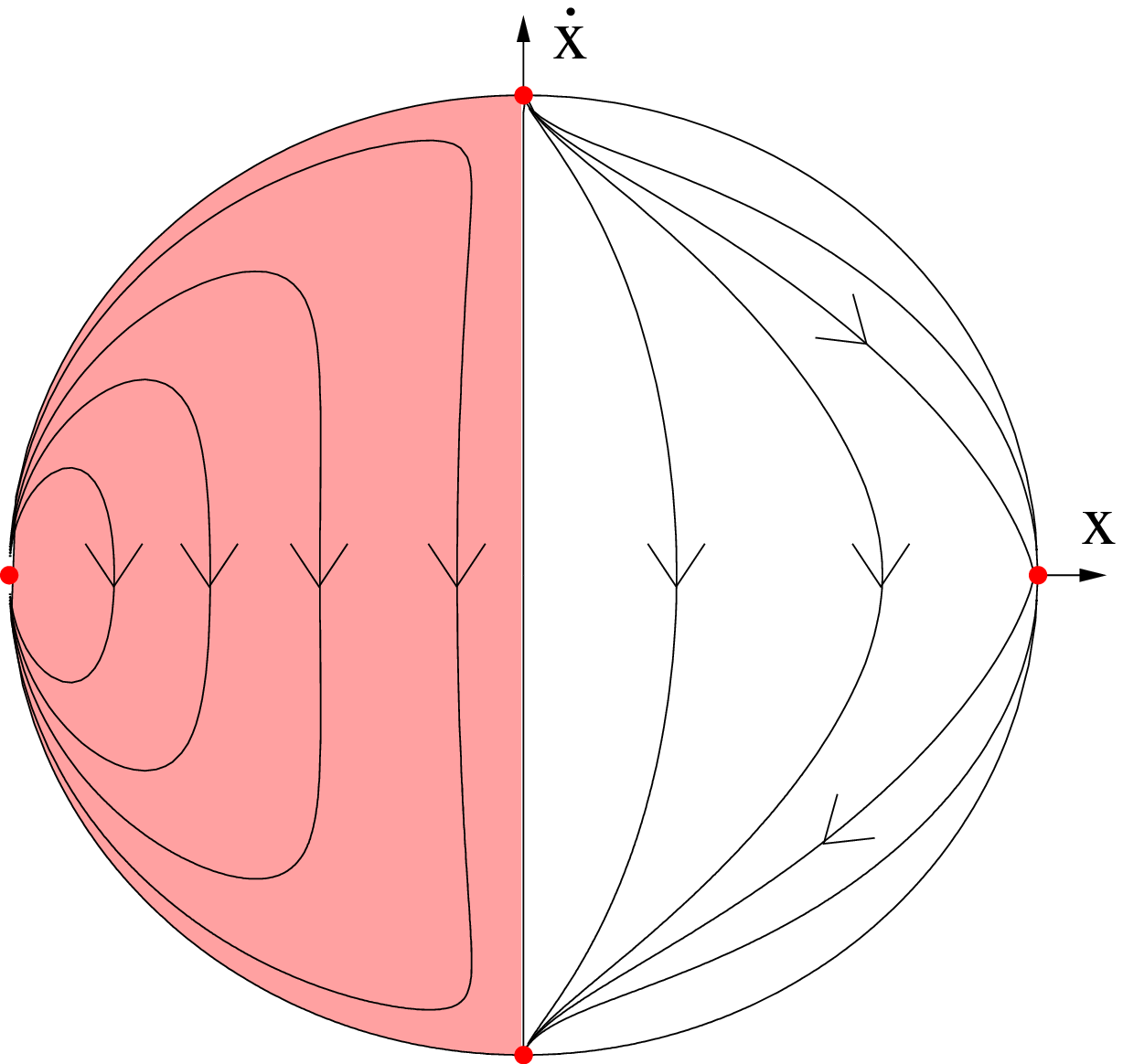} \\ [0.0cm]
\end{array}$
\end{center}
\caption{The phase portraits for the Cardassian potentials drawn in Fig.~\ref{fig3}.
All representative phase portraits for different choice of pair ($w,n$) are shown. The observation of SNIa 
favors the cases {\bf (a),(b),(c)} for which the static critical points are present. In the neighborhood of 
this solution there are trajectories with very slow expansion or loitering \cite{sahni92}. In the case 
{\bf(d)}, there is a complex critical point at ($0,0$) and for the cases {\bf (e),(f)} there is no critical 
points at a finite domain. The phase portraits {\bf (a)}, {\bf (b)} and {\bf (c)} are topologically 
equivalent. The critical point at infinity ($x = \infty,\dot{x} = \infty$) is a stable node for expanding 
models. It represents a big-rip singularity in the future while $x = 0$, $\dot{x} = 0$ represents 
the initial singularity which is dominated by the standard matter term. Note that models with a big-rip 
singularity in the future are typical because expanding models have a global attractor at the circle at 
infinity.}
\label{fig4}
\end{figure}

Now, let us concentrate on the models for which the potential function is upper convex. As we demonstrate 
in our previous papers (\cite{szydlowski03,szydlowski03a,szydlowski03c}) such a form is favored by SNIa data
(see also next section with reconstruction of $V(a)$  from Knop's sample of SNIa data). 
It is interesting that classification of all possible (representative) evolutional paths in the phase
plane can be simply performed in terms of a pair ($w,n$) (see Fig~\ref{fig2}).
There are presented domains (shaded regions) on the plane $(w,n)$ for which the Cardassian models, 
described by the potential function (\ref{eq:9a}), start to accelerate for $a>0$ in Fig.~\ref{fig2}.
In these cases the potential function $V(x)$ of the system has a maximum for $a>0$ and is upper convex.
The phase portraits for the representative cases of Cardassian potentials, parameterized by $w$ and $n$, 
signed as ${\bf (a)-(f)}$ in Fig.~\ref{fig2} and drawn in Fig.~\ref{fig3}, have been shown in 
Fig.~\ref{fig4}.

Main advantage of this reduction procedure is the application of dynamical system approach to the case
of multi-fluids and in general to any fluid for which energy density can be expressed as a function 
of scale factor. For illustration let us consider a universe filled with matter and radiation. Then
$\rho=\rho_{m,0}a^{-3}+\rho_{r,0}a^{-4}$ and an additional term named the Cardassian term can be 
interpreted as some fictitious fluid with energy density $\rho_{\rm DE}$ for which the equation of 
state is $p_{\rm DE}=w_{\rm DE}\rho_{\rm DE}$, where
\begin{equation*}
w_{\rm DE} = (n-1)+\frac{1}{1+a\Omega_{m,0}/\Omega_{r,0}}.
\end{equation*}
In this case $\rho_{\rm eff}=\rho+\rho_{\rm DE}$ and the potential function for the corresponding
Hamiltonian system takes the exact form (see Table~\ref{tab1}). 

Let us now consider a class of the so-called generalized Cardassian models, given in the form
\begin{equation}
\frac{{\dot{x}}^{2}}{x^{2}} = g(x), 
\label{eq:9d}
\end{equation}
where $x$ is a dimensionless scale factor expressed in terms of the present value of the scale factor 
$a_{0}$; $x = a/a_{0}$; dot denotes here differentiation with respect to the new time variable $\eta$
such that $dt/H_{0} = d \eta$.

Note that function $g(x)$ plays the role of effective energy density in standard Cardassian cosmology.
We rewrite the generalized Friedmann equation to the form
\begin{equation}
\mathcal{E} = \frac{{\dot{x}}^{2}}{2} - g(x)\frac{x^{2}}{2} \equiv 0.
\label{eq:9e}
\end{equation}
Equation (\ref{eq:9e}) is an energy function, while  $-g(x) x^{2}/2$ plays the role of potential 
energy for particle-universe of unit mass moving in one-dimensional potential.

Differentiation of (\ref{eq:9e}) with respect to $\eta$ gives the Newtonian equation of motion in the form
\begin{equation}
\ddot{x} = - \frac{\partial V}{\partial x}\quad {\rm and} \quad V(x) = - g(x)\frac{x^{2}}{2}.
\label{eq:9f}
\end{equation}

The Hamiltonian for our problem takes the following form 
\begin{equation}
\mathcal{H} = \frac{1}{2} {P_{x}}^{2} + V(x),
\label{eq:9g}
\end{equation}
where $P_{x} = \dot{x}$.

Equation (\ref{eq:9f}) can be reduced to the form of system of second order differential equation
\begin{align}
\dot{x} &= y,\nonumber \\
\dot{y} &= - \frac{\partial V}{\partial x}.
\label{eq:9h}
\end{align}
with first integral $\dot{x}^{2}/2+V(x) \equiv 0$.

It is interesting that all knowledge that is needed about the stationary solution (critical points)
and its stability can be simply deduced from the geometry of the potential function.

\begin{table}[!ht]
\caption{The Cardassian models}
\begin{center}
\begin{ruledtabular}
\tiny
\begin{tabular}{c|l|c}
The model & Form of the potential function & References \\
\hline
\parbox{3cm}{Cardassian model with both dust and radiation} &
$V(x)=-\frac{1}{2}\Omega_{m,0}x^{-2}\Bigg[\bigg(x+\frac{\Omega_{r,0}}{\Omega_{m,0}}\bigg)+x^{-4(n-1)}\frac{1-\Omega_{r,0}-\Omega_{m,0}}{\Omega_{m,0}}\bigg(\frac{x+\Omega_{r,0}/\Omega_{m,0}}{1+\Omega_{r,0}/\Omega_{m,0}}\bigg)^{n}\Bigg]$ &
\cite{sen03} \\
\hline
\parbox{3cm}{Generalized Chaplygin gas as a Cardassian model} &
$V(x)=-\frac{1}{2}\Bigg[\Omega_{m,0}x^{-1}+\Omega_{{\rm Chap},0}x^{2}\bigg(A_{s}+\frac{1-A_{s}}{x^{3(1+\alpha)}}\bigg)^{1/(1+\alpha)}\Bigg]$ &
\cite{kamenshchik01,bento02} \\
\hline
\parbox{3cm}{Cosmology arises from MOND gravity} &
$V(x)=-\frac{1}{2}\Bigg[\Omega_{m,0}x^{-1}+\Omega_{{\rm MOND},0}\log{x}+(1-\Omega_{m,0})\Bigg]$ &
\cite{lue03,lue03a} \\ 
\hline
\end{tabular}
\end{ruledtabular}
\end{center}
\label{tab1}
\end{table}

\begin{figure}[!ht]
\begin{center}
$\begin{array}{c@{\hspace{0.2in}}c}
\multicolumn{1}{l}{\mbox{\bf (a)}} & 
\multicolumn{1}{l}{\mbox{\bf (b)}} \\ [-0.5cm]
\includegraphics[scale=0.32,angle=270]{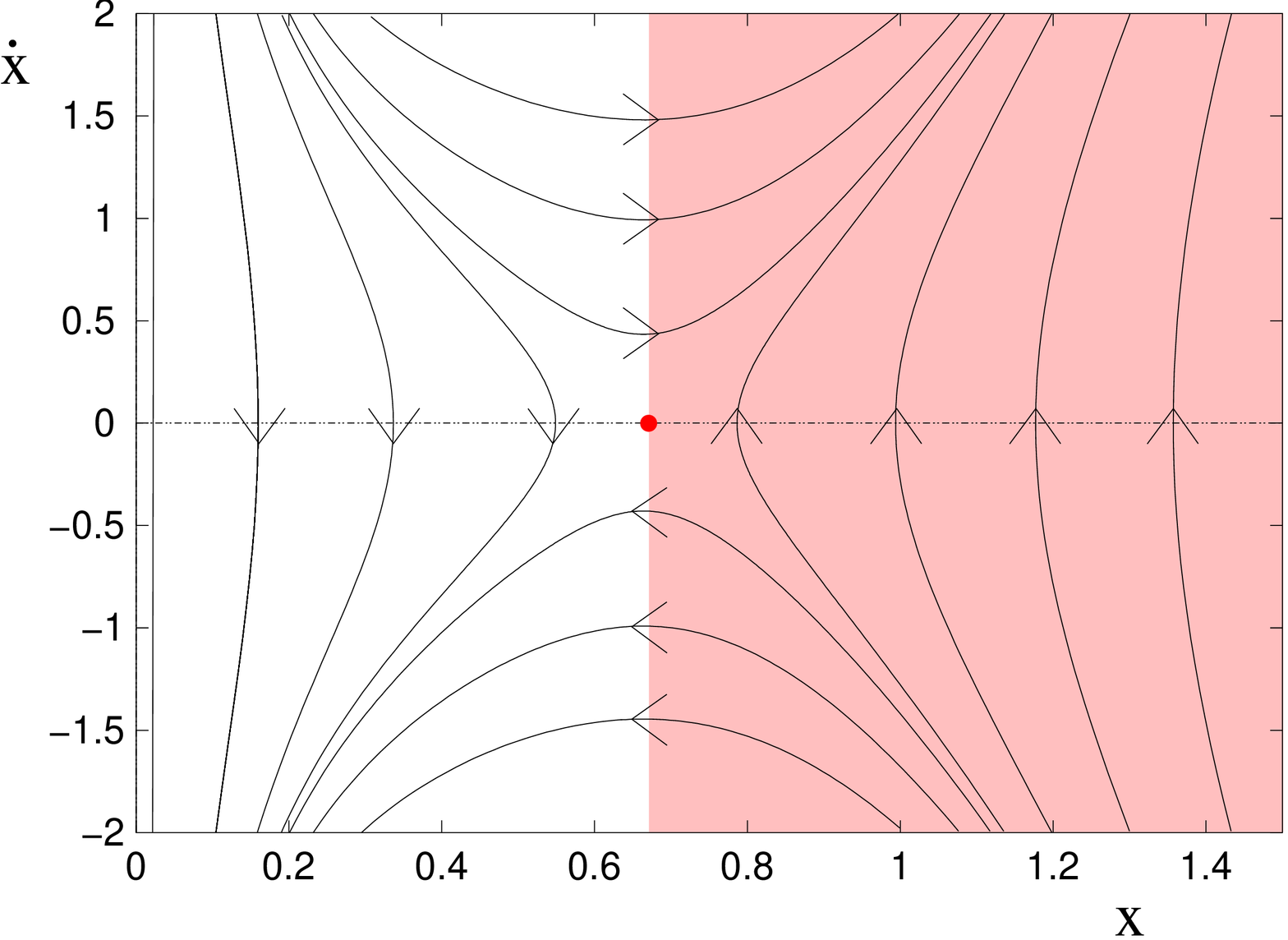} & 
\includegraphics[scale=0.32,angle=270]{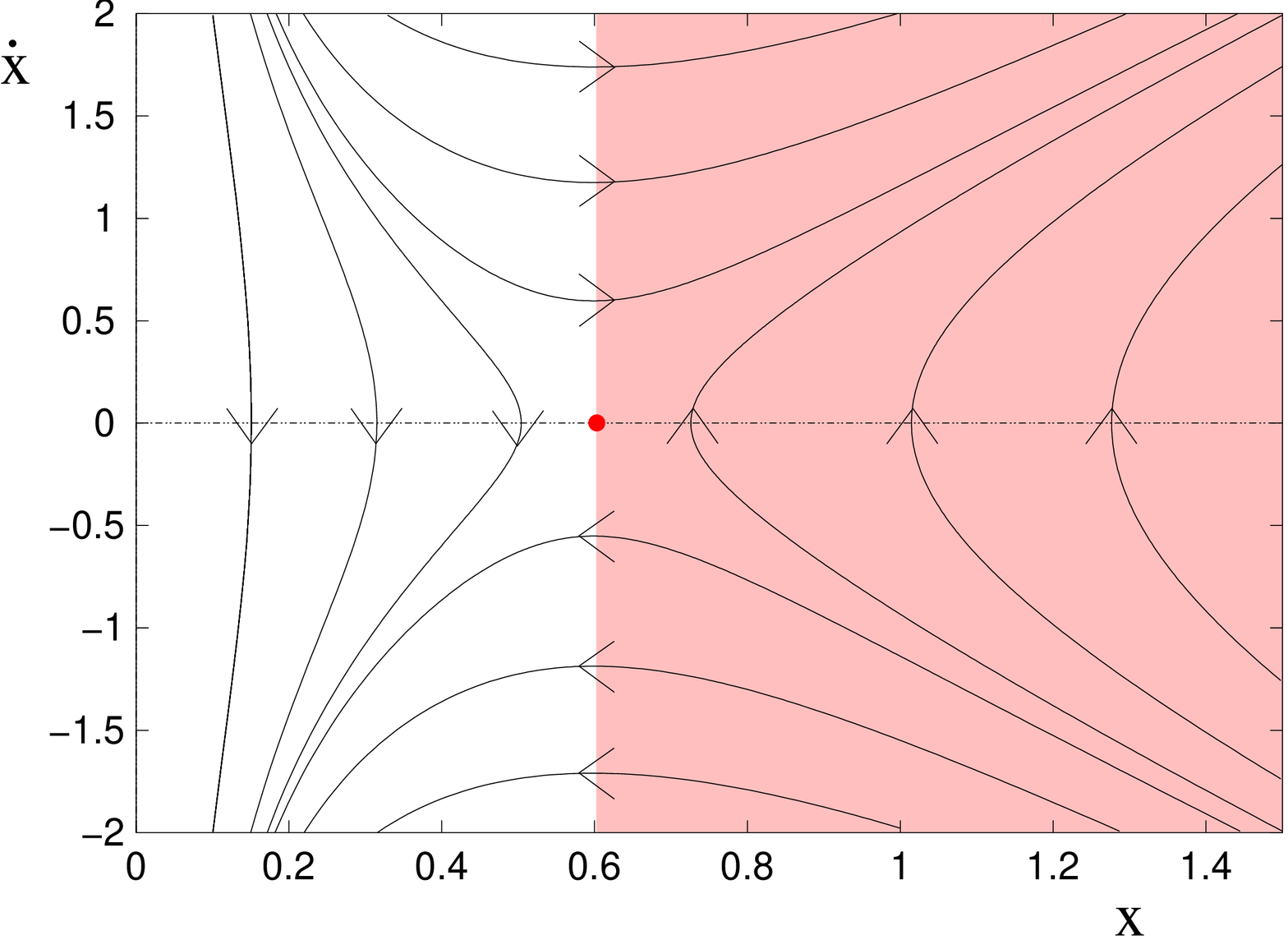} \\ [0.4cm]
\multicolumn{1}{l}{\mbox{\bf (c)}} & 
\multicolumn{1}{l}{\mbox{\bf (d)}} \\ [-0.5cm]
\includegraphics[scale=0.32,angle=270]{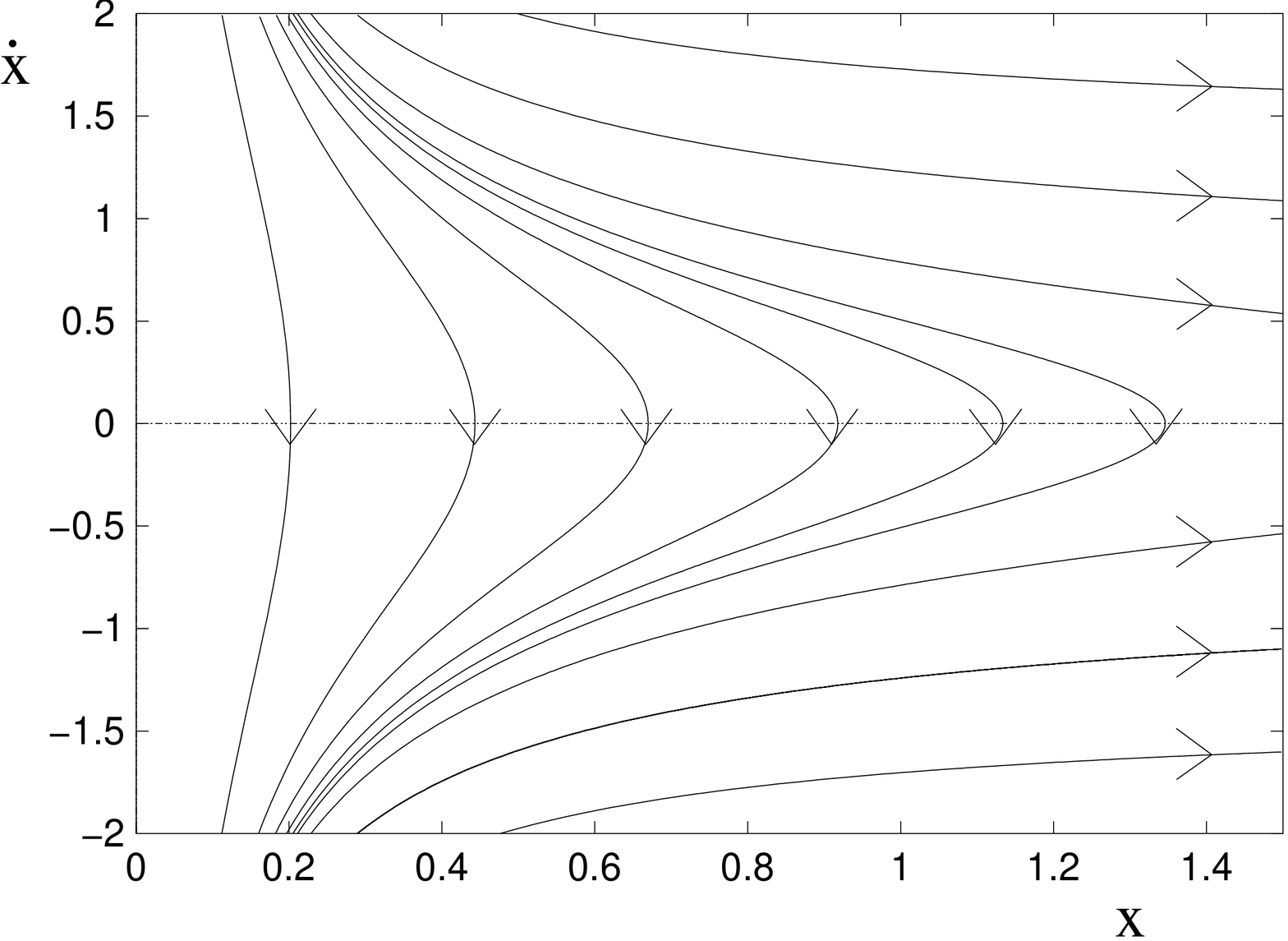} & 
\includegraphics[scale=0.32,angle=270]{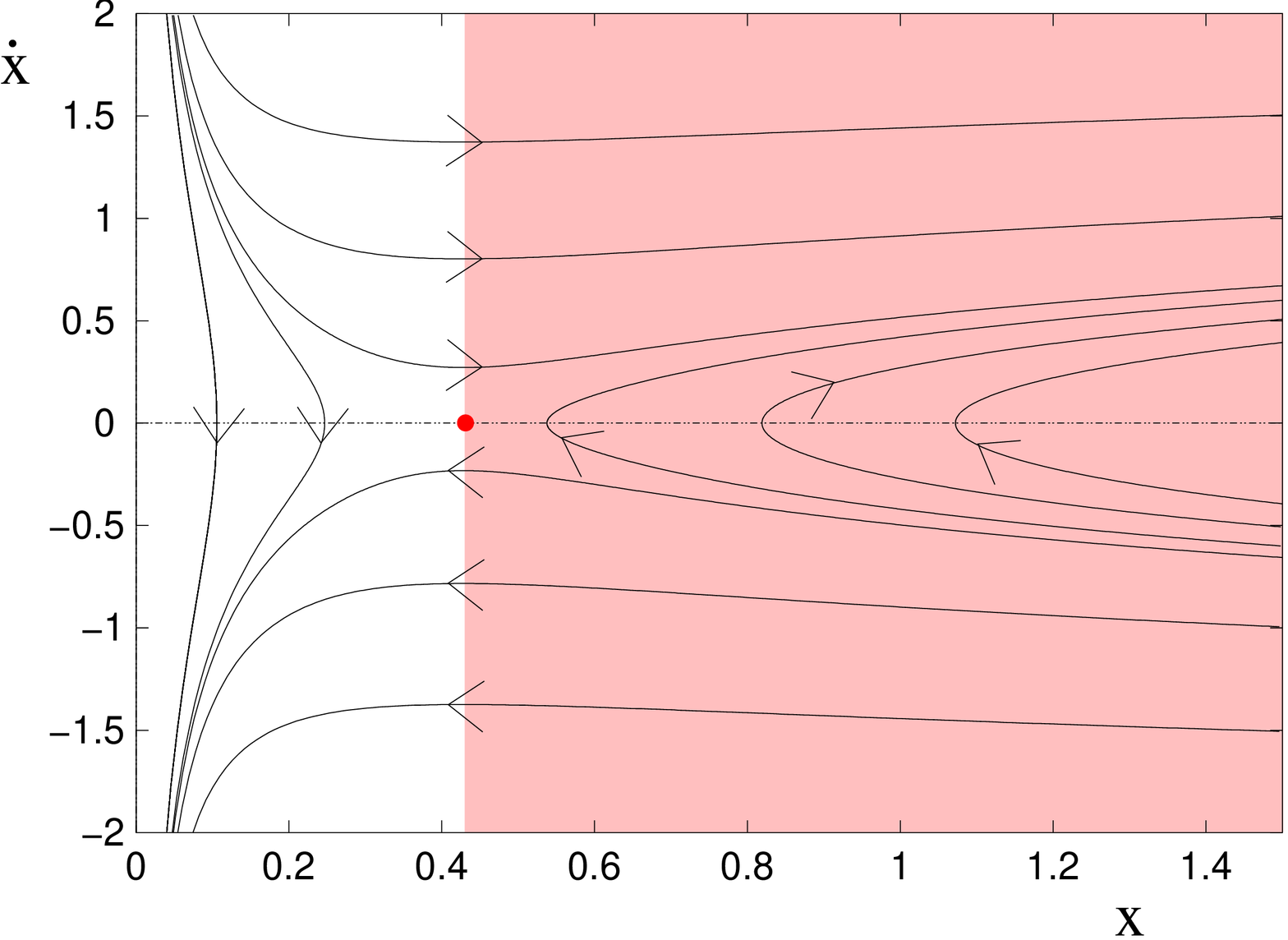} \\ [0.4cm]
\end{array}$
\end{center}
\caption{The phase portraits for the Cardassian potential functions from Table~\ref{tab1}. 
{\bf (a), (b), (c)}--Cardassian models with dust and radiation for $n=-1,0,1$ respectively, 
{\bf (d)}--MOND model. The shaded region is the region of accelerated expansion of the universe. 
The systems {\bf (a), (b), (c)} are structurally stable because of the presence of saddle points 
in a finite region.}
\label{fig5}
\end{figure}

We can find how different solutions are distributed on the phase plane. Whether they are generic 
(typical) or exceptional (of zero measure). In terms of dynamical system notions the 
question is whether they constitute a global attractor or a separatrix. It is interesting to check, for 
example, under which conditions as an asymptotic state for late time we obtain de Sitter solution-exit for 
inflation (acceleration). We can prove that, if the potential function for large $x$ 
($x \rightarrow \infty$) is negative and it is a homogeneous function (of its argument) of degree two, 
then as a critical point we obtain stable critical point at infinity. To prove this it is sufficiently to 
consider the system in projective coordinates $z = 1/x$, $u = y/x$. Hence we obtain the interesting critical
point $z=0$, $u=u_{0}$ (it represents de Sitter solution) if only $V(z)$ satisfy the condition
$dV/dz=2/zV(z)$. Therefore $V(z) \propto z^{2}$. From the linearization matrix and the
characteristic equation we find the eigenvalues of the linearization matrix $\lambda_{1}=-u_{0}$, 
$\lambda_{2}=-2u_{0}$. They are negative for expanding models. Because both eigenvalues are real
and negative, the critical point is a stable node. It is located on the intersection of a circle at 
infinity $z=0$ $- \infty < u < + \infty$ and the trajectory of the flat model $u^{2}/2z^{2}+V(z)=0$. 

The potential function for different generalizations of Cardassian models are shown in Table~\ref{tab1}.

\section{Exact solution and duality relation}

For our investigations of exact solutions and transformation relations between FRW model solutions
with the sub-negative equation of state ($1 + \gamma > 0$, $p = \gamma \rho$, $\gamma = {\rm const.}$) and such 
models with super-negative equation of state ($\gamma < -1$), it can be useful to rewrite the basic dynamical 
equation to the form
\begin{equation}
\Bigg(\frac{\dot{x}}{x}\Bigg)^{2}=\Omega_{i,0}x^{-3(1+\gamma_{i})}+(1-\Omega_{i,0})x^{-3n(1+\gamma_{i})},
\label{eq:10}
\end{equation}
where $x \equiv a/a_{0}$ is the scale factor in the units of its present value $a_{0}$; 
$\Omega_{i,0}=\rho_{i,0}/(3H_{0}^{2})$ and $\Omega_{{\rm Card},0}=1-\Omega_{i,0}$ are the density parameters
for matter which satisfy the equation of state $p=\gamma_{i} \rho$ ($\gamma={\rm const.}$) and for some
fictitious fluid which mimic the Cardassian term of $\rho^{n}$ type; $\Omega_{k,0}=0$ for simplicity.

The equation of state of the fictitious fluid which reproduces the Cardassian term is
\begin{equation}
p_{\rm Card}=[n(1+\gamma)-1]\rho.
\label{eq:11}
\end{equation}
Therefore, if $n(1+\gamma)<0$, then an additional noninteracting fluid possesses property of super-negative 
pressure (i.e. phantom matter). The generic property of phantom cosmology is the existence of a new type 
singularities, since they appear for infinite values of the scale factor. Hence, if we consider a standard 
matter (which doesn't violate weak energy condition) then it is sufficiently that $n<0$ for the appearance 
of big-rip type singularity. In this case the Cardassian term (fluid) dominates for a large $x$ 
($a \gg a_{0}$) while the material term is important near the initial singularity. If the Cardassian term 
dominates the material term we obtain the following asymptotic
\begin{equation}
x(t) \propto (t-t_{0})^{\frac{2}{3n(1+\gamma)}}.
\label{eq:12}
\end{equation}
From (\ref{eq:12}) it is clear that the big-rip singularity appears in a finite time $t_{0}$ for Cardassian
models if only (a) $n<0$ and $(1+\gamma)>0$ or (b) $n>0$ and $(1+\gamma)<0$. In the last case 
$x(t) \propto (t-t_{0})^{2/[3(1+\gamma)]}$ for large $x$ and Cardassian term is negligible (pure phantom
cosmology approximation is obvious).

To find the exact solutions of (\ref{eq:10}) it is useful to introduce a new time variable $\eta$ instead
of original one $t$, namely
\begin{equation}
t \rightarrow \eta \colon x^{-\frac{3(1+\gamma)}{2}}dt=d\eta.
\label{eq:13}
\end{equation}
Of course the new time variable is a monotonic function of the original time and exact solutions can be 
expressed in terms of $\eta$ in the following way
\begin{equation}
x(\eta)=\bigg(\frac{\Omega_{i,0}}{1-\Omega_{i,0}}\bigg)^{\beta}\Bigg[\sinh{\bigg(\frac{\eta-\eta_{0}}{2\beta}\bigg)}\sqrt{\Omega_{i,0}}\Bigg]^{-2\beta},
\label{eq:14}
\end{equation}
where $\beta=[3(1+\gamma)(1+n)]^{-1}$.

In the special cases of domination of the Cardassian contribution and matter we obtain the following 
asymptotic solutions near the big-rip singularities
\begin{align}
\Omega_{i,0}&=0, \quad \Omega_{{\rm Card},0}=1, \nonumber \\
x(\eta)& \propto \exp{(\eta-\eta_{0})}, \quad x(t) \propto (t-t_{0})^{\frac{2}{3n(1+\gamma)}}
\label{eq:15}
\end{align}
and
\begin{align} 
\Omega_{i,0}&=1, \quad \Omega_{{\rm Card},0}=0, \nonumber \\
x(\eta)& \propto \exp{(\eta-\eta_{0})}, \quad x(t) \propto (t-t_{0})^{\frac{2}{3(1+\gamma)}}.
\label{eq:16}
\end{align}
Let us note that in time $\eta$ the big-rip singularity is moved to infinity. Finally, we can conclude
that analogical type of unwanted future singularity when infinite energy density is reached during a finite
cosmological time appear in Cardassian scenario. Moreover it can appear without assumption of super-negative 
pressure for matter. The existence of this kind of singularity is a generic property of the Cardassian models
when $n<0$.

It is easy to notice that equation (\ref{eq:10}), as well as relation (\ref{eq:13}), preserves its form
under the change both position variables and sign of equation of state parameter $(1+\gamma)$
\begin{equation}
x \rightarrow \frac{1}{x} \quad {\rm and} \quad (1+\gamma) \rightarrow -(1+\gamma).
\label{eq:17}
\end{equation}
From this observation we obtain a simple rule which gives us a solution of the Cardassian models in 
phantom's epochs if we know a solution in quintessence epoch with sub-negative equation of state. 
If $x_{\gamma}$ is the solution of (\ref{eq:10}) then $x^{-1}_{-(\gamma+2)}$ is also the solution of 
(\ref{eq:10}). In other words, taking $\gamma=-2/3$ like for topological defects we immediately obtain as a 
solution $x^{-1}$ for phantoms ($\gamma=-4/3$).

Let us now briefly comment this property. The existence of symmetry relation (\ref{eq:17}) between the 
phantom models and quintessence one indicates the presence of duality in the dynamical behavior of the 
phantom and the quintessence field alike to the scale factor duality symmetry in super-string theory 
\cite{meissner91,dabrowski03}.

\section{Reconstruction of the dynamics for the Cardassian model from the sample of Knop's supernovae data}

It was shown in our previous papers \cite{szydlowski03,szydlowski03c} that the potential function $V[x(z)]$,
which completely determines the dynamics of the Cardassian model (see section II B), can be simply 
reconstructed from the observed magnitude-redshift relation of distant supernovae.

As it is well known the relation between the apparent magnitude $m$ and the dimensionless luminosity distance 
$\mathcal{D}_{L}$ is
\begin{equation}
m=\mathcal{M}+5\log{[\mathcal{D}_{L}(\Omega_{m,0},\Omega_{{\rm Card},0},n,z)]},
\label{erec:1}
\end{equation}
where $\mathcal{M}=M-5\log{H_{0}}+25$; $M$ is the absolute magnitude; $H_{0}$ is the Hubble 
constant and the dimensionless luminosity distance for the flat model takes the form
\begin{gather}
\mathcal{D}_{L}(\Omega_{m,0},\Omega_{{\rm Card},0},n,z)=H_{0}d_{L}(\Omega_{m,0},\Omega_{{\rm Card},0},n,H_{0},z) \nonumber \\
=H_{0}(1+z)\int_{0}^{z}\frac{dz'}{H(\Omega_{m,0},\Omega_{{\rm Card},0},n,H_{0},z')} 
=(1+z)\int_{0}^{z}\frac{dz'}{\sqrt{\Omega_{m,0}(1+z')+\Omega_{{\rm Card},0}(1+z')^{3n-2}}}.
\label{erec:2}
\end{gather}
Due to the existence of such a relation it is possible to calculate the potential function $V$ which is
\begin{gather}
V[a(z),\Omega_{m,0},\Omega_{{\rm Card},0},n,H_{0},z]=-\frac{1}{2}H^{2}(\Omega_{m,0},\Omega_{{\rm Card},0},n,H_{0},z)a^{2}(z)
\nonumber \\=-\frac{1}{2}\Bigg[\frac{d}{dz}\Bigg(\frac{d_{L}(\Omega_{m,0},\Omega_{{\rm Card},0},n,H_{0},z)}{1+z}\Bigg)\Bigg]a^{2}(z)
=-\frac{1}{2}H_{0}^{2}\Big[\Omega_{m,0}(1+z)+\Omega_{{\rm Card},0}(1+z)^{3n-2}\Big].
\label{erec:3}
\end{gather}
Relation (\ref{erec:3}) can be rewritten in the following (dimensionless) form
\begin{equation}
V[x(z),\Omega_{m,0},\Omega_{{\rm Card},0},n]=-\frac{1}{2}\Big[\Omega_{m,0}x^{-1}+\Omega_{{\rm Card},0}x^{-3n+2}\Big].
\label{erec:4}
\end{equation}

We determine the model parameters ($\Omega_{m,0}$, $\Omega_{{\rm Card},0}$, $n$, $\mathcal{M}$) as well as
the potential function $V$ using a $\chi^{2}$ minimization procedure, where $\chi^{2}$ is described by the
formula
\begin{equation}
\chi^{2}=\sum_{j}\frac{[m_{t,j}(\Omega_{m,0},\Omega_{{\rm Card},0},n,\mathcal{M})-m_{o,j}]^{2}}{\sigma_{m,j}^{2}},
\label{erec:5}
\end{equation}
where $m_{o,j}$ is the measured magnitude; $m_{t,j}$ can be calculated from the equations (\ref{erec:1}) 
and (\ref{erec:2}) for given values of parameters $\Omega_{m,0}$, $\Omega_{{\rm Card},0}$, $n$, 
$\mathcal{M}$ and $z$; $\sigma_{m,j}$ is the magnitude measurement uncertainty. The summation in the equation
(\ref{erec:5}) is over all of the observed supernovae.

\begin{table}[!ht]
\linespread{0.5}
\caption{Results of the statistical analysis for the Cardassian model from distant type Ia 
supernovae data (Subset 1 of 58 supernovae compiled by Knop et al. \cite{knop03}), 
obtained from the best fit method ($\chi^{2}$ minimization) and from the likelihood method 
(denoted with max($\mathcal{L}$)). In the case {\bf (a)} the analysis was prepared with fixed $\mathcal{M}$ (chosen as the 
best-fitted value of $\mathcal{M}$ for the model). The probability density distributions 
(from the likelihood procedure) for the estimated cosmological parameters are shown in Fig.~\ref{stata2}.}
\begin{center}
\begin{ruledtabular}
\begin{tabular}{c|c|c|c|c|c|c|c}
  & $\Omega_{m,0}$ & $\Omega_{{\rm Card},0}$ & $n$ & $\mathcal{M}$ & $\chi^{2}$ & $z_{max}$ & $V_{max}$ \\
\hline
best fit & $0.55$ & $0.45$ & $-4.46$ & $-3.62$ & $53.42$ & $0.189$ & $-1492.96$ \\
max($\mathcal{L}$) & $0.57 ^{+0.05}_{-0.05}$ & $0.43 ^{+0.05}_{-0.05}$ & $-4.36 ^{+1.69}_{-2.17}$ & $-3.62$&&&\\
\hline
best fit & $0.55$ & $0.45$ & $-4.46$ & $-3.62$ & $53.42$ & $0.189$ & $-1492.96$ \\
max($\mathcal{L}$) & $0.54 ^{+0.07}_{-0.06}$ & $0.46 ^{+0.06}_{-0.07}$ & $-4.14 ^{+3.55}_{-4.31}$ & $-3.61 ^{+0.08}_{-0.11}$&&&\\
\end{tabular}
\end{ruledtabular}
\end{center}
\label{results1}
\end{table}
\begin{figure}[!ht]
\begin{center}
$\begin{array}{c@{\hspace{0.2in}}c@{\hspace{0.2in}}c}
\multicolumn{1}{l}{\mbox{\bf (a)}} & 
\multicolumn{1}{l}{\mbox{\bf (b)}} \\ [0.2cm]
\includegraphics[scale=0.45, angle=0]{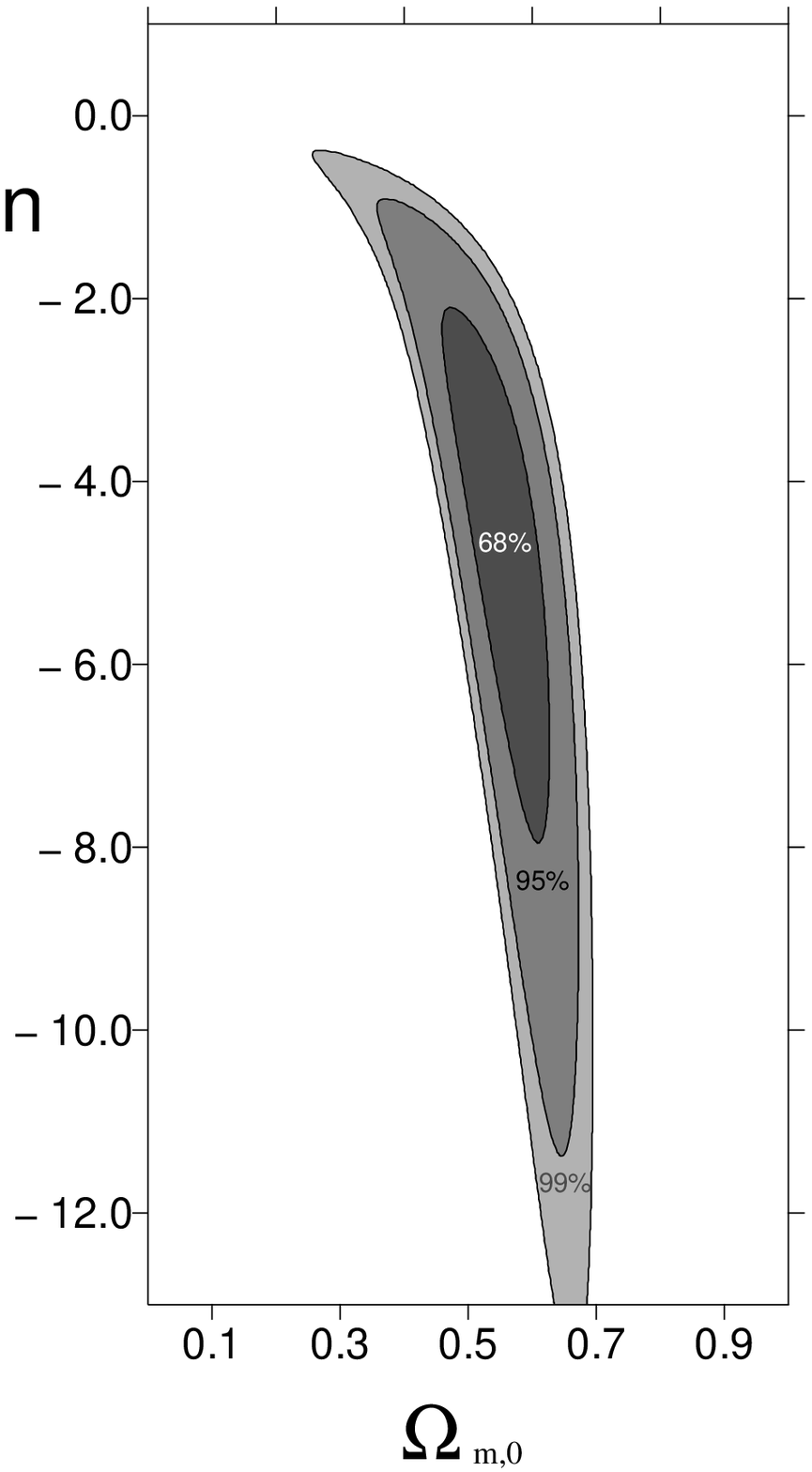} & 
\includegraphics[scale=0.45, angle=0]{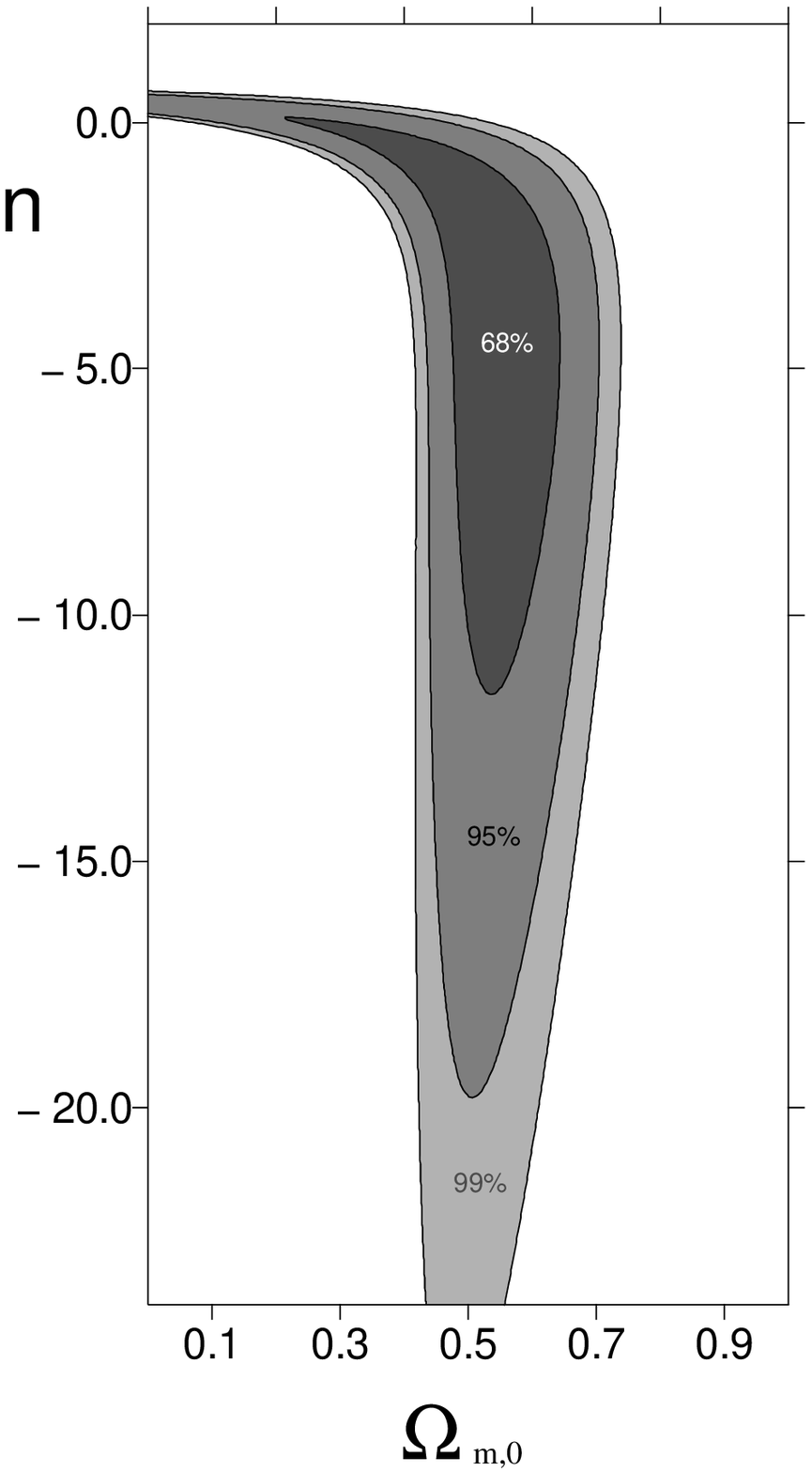} \\ [0.0cm]
\end{array}$
\end{center}
\linespread{0.5}
\caption{Confidence levels on the plane $(\Omega_{m,0},n)$ for the Cardassian model obtained by the integration of
the likelihood function $\mathcal{L}(\Omega_{m,0},\Omega_{{\rm Card},0},n)$ over {\bf (a)} $\Omega_{m,0}$, 
$\mathcal{M}$ was chosen as the best-fitted value of $\mathcal{M}$ for the model in minimization procedure;
{\bf (b)} $\Omega_{m,0}$ and $\mathcal{M}$  for the subset 1 of 58 supernovae compiled by Knop et al. 
\cite{knop03}. $\mathcal{M}$ was chosen as the best-fitted
(see Table \ref{results1}).}
\label{stata1}
\end{figure}
\begin{figure}[!ht]
\begin{center}
$\begin{array}{c@{\hspace{0.2in}}c}
\multicolumn{1}{l}{\mbox{\bf (a)}} & 
\multicolumn{1}{l}{\mbox{\bf (b)}} \\ [-0.5cm]
\includegraphics[scale=0.24, angle=270]{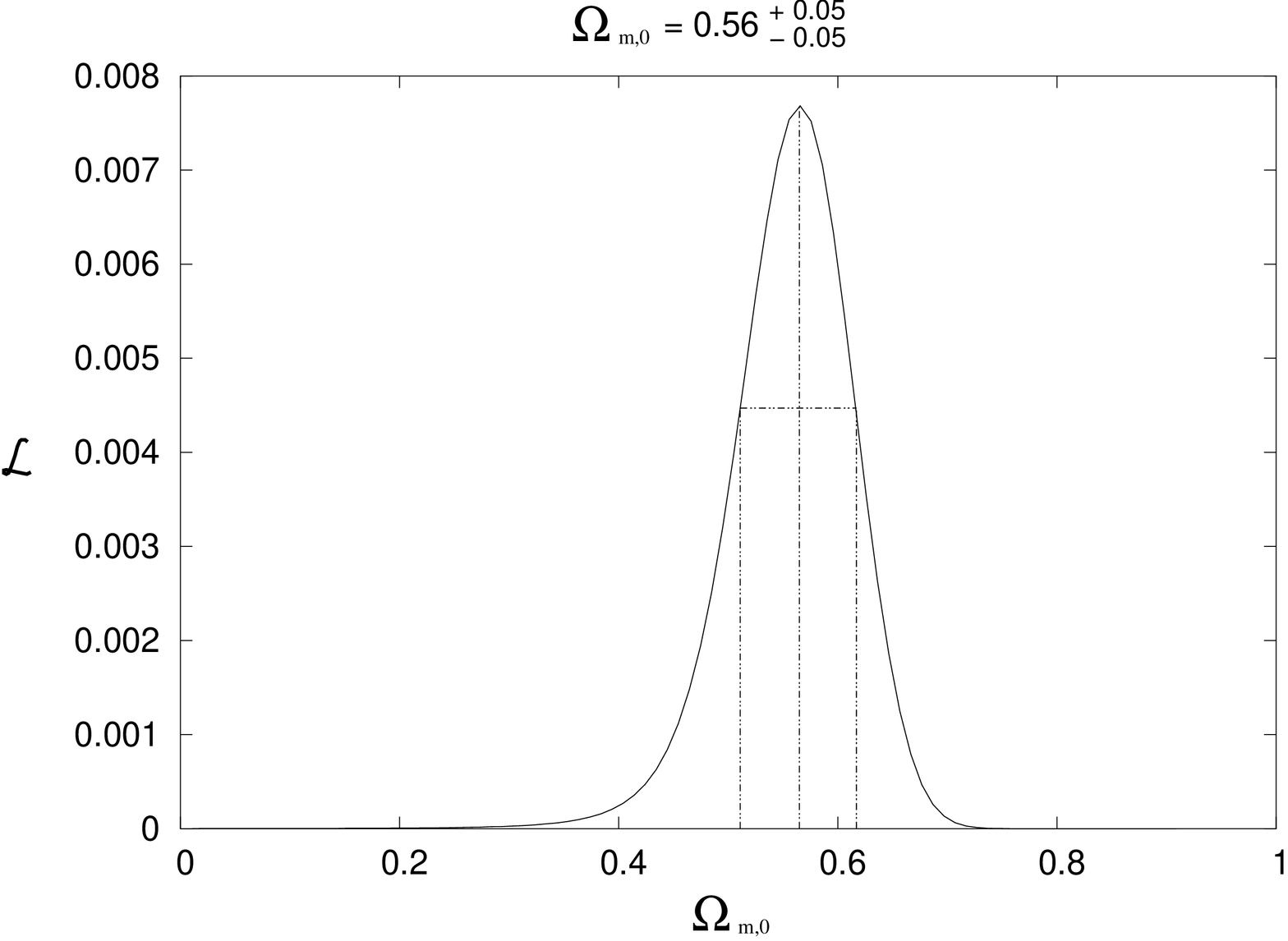} & 
\includegraphics[scale=0.24, angle=270]{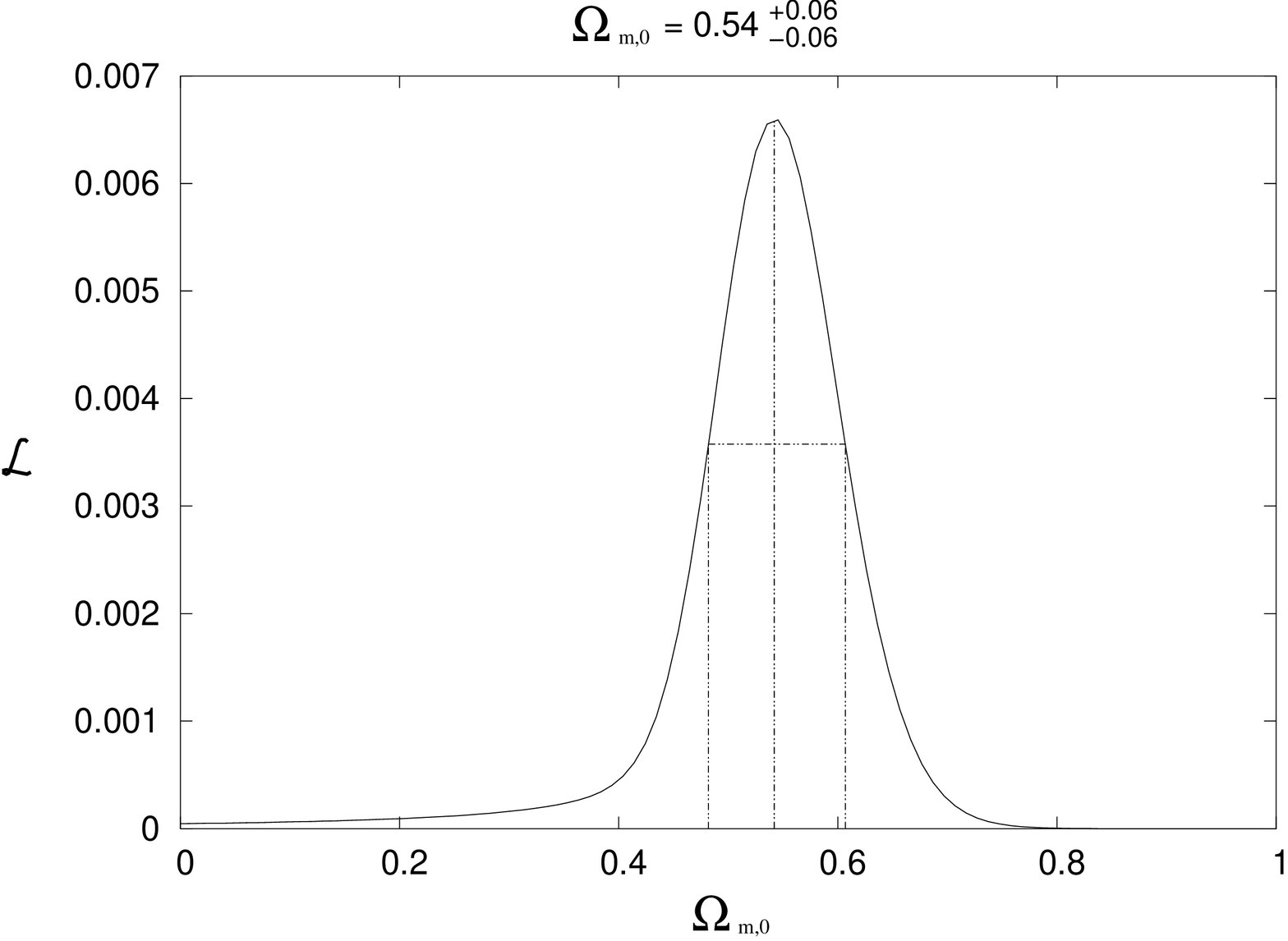} \\ [0.4cm]
\multicolumn{1}{l}{\mbox{\bf (c)}} & 
\multicolumn{1}{l}{\mbox{\bf (d)}} \\ [-0.5cm]
\includegraphics[scale=0.24, angle=270]{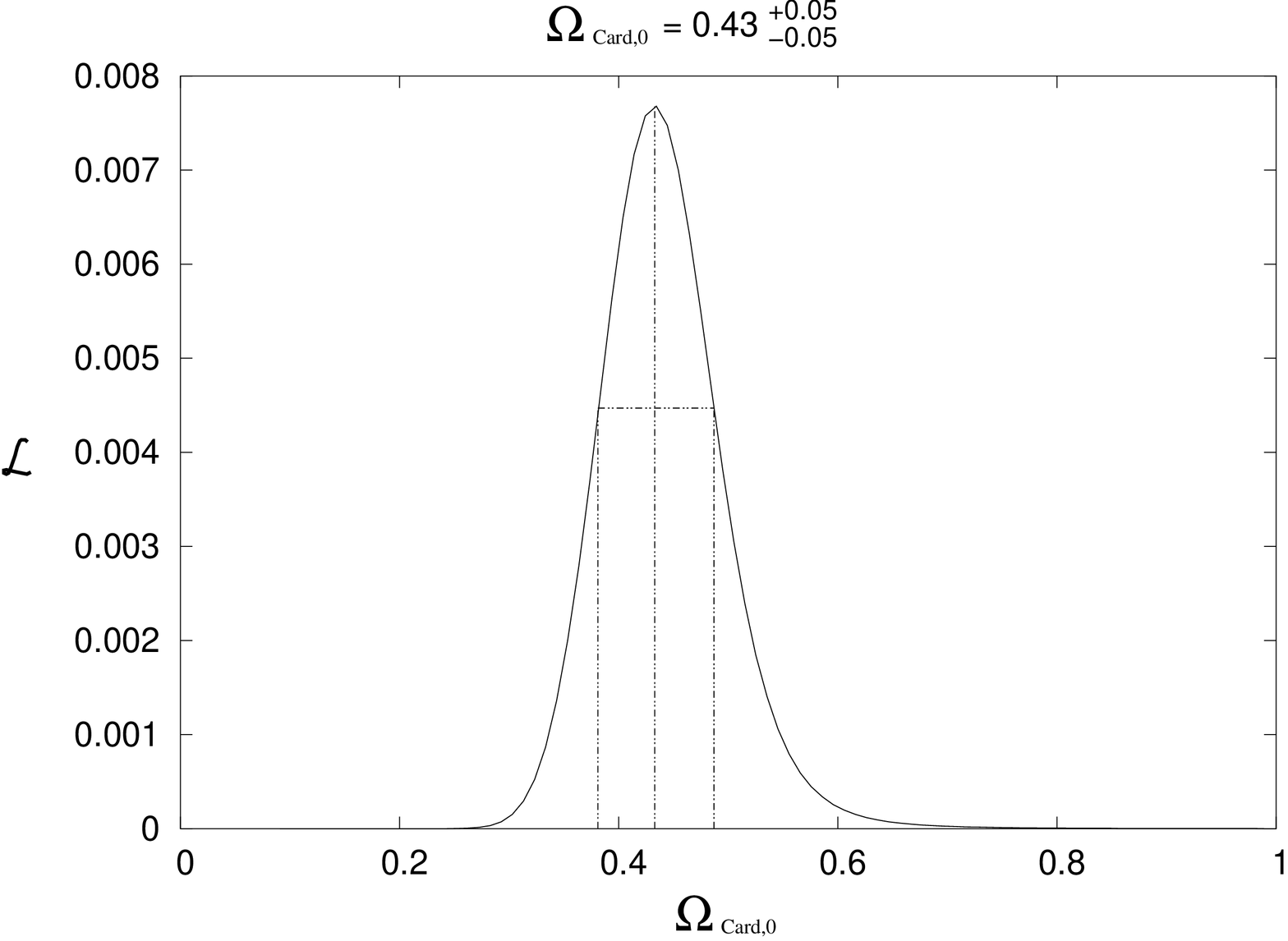} & 
\includegraphics[scale=0.24, angle=270]{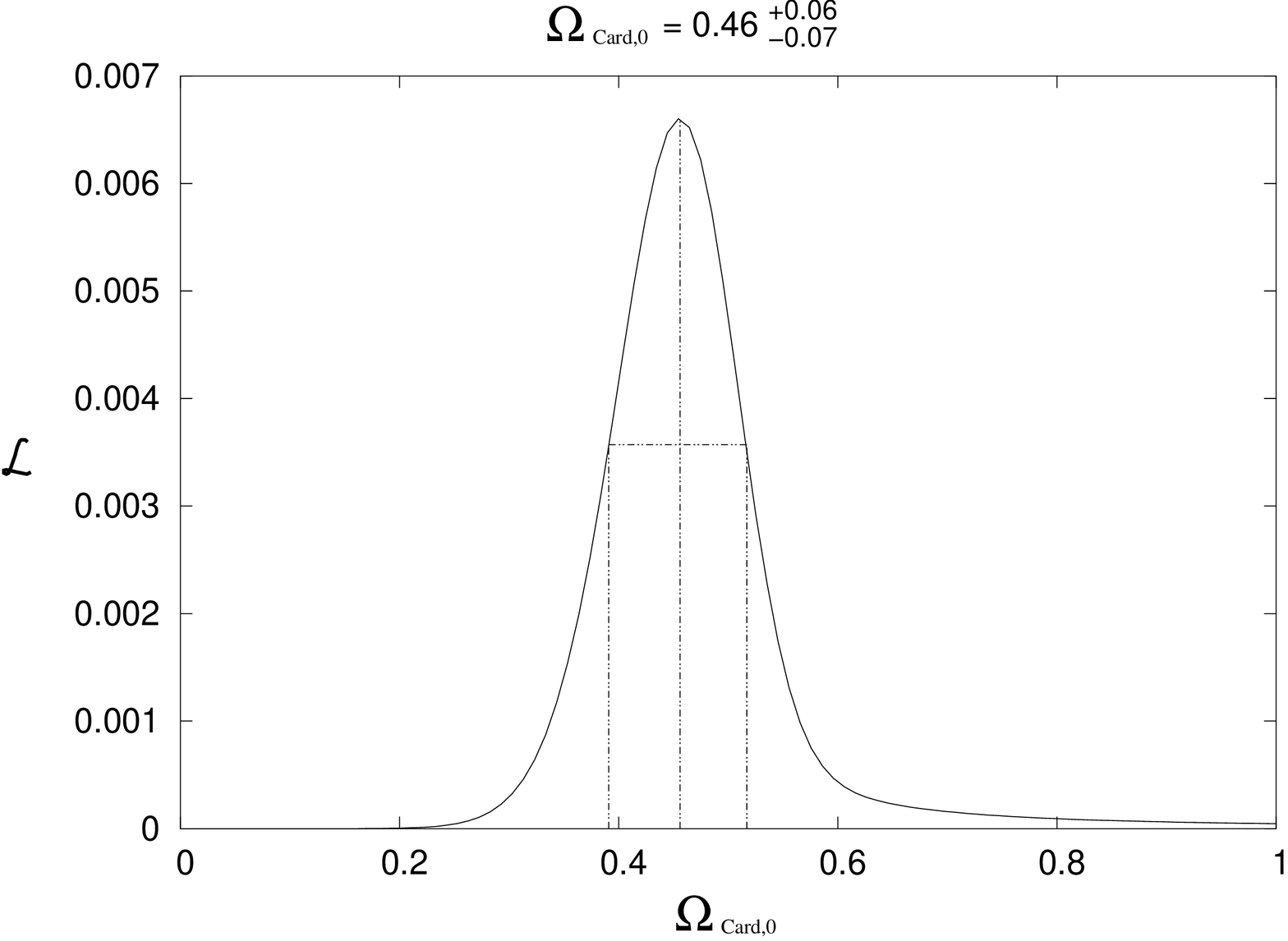} \\ [0.4cm]
\multicolumn{1}{l}{\mbox{\bf (e)}} & 
\multicolumn{1}{l}{\mbox{\bf (f)}} \\ [-0.5cm]
\includegraphics[scale=0.24, angle=270]{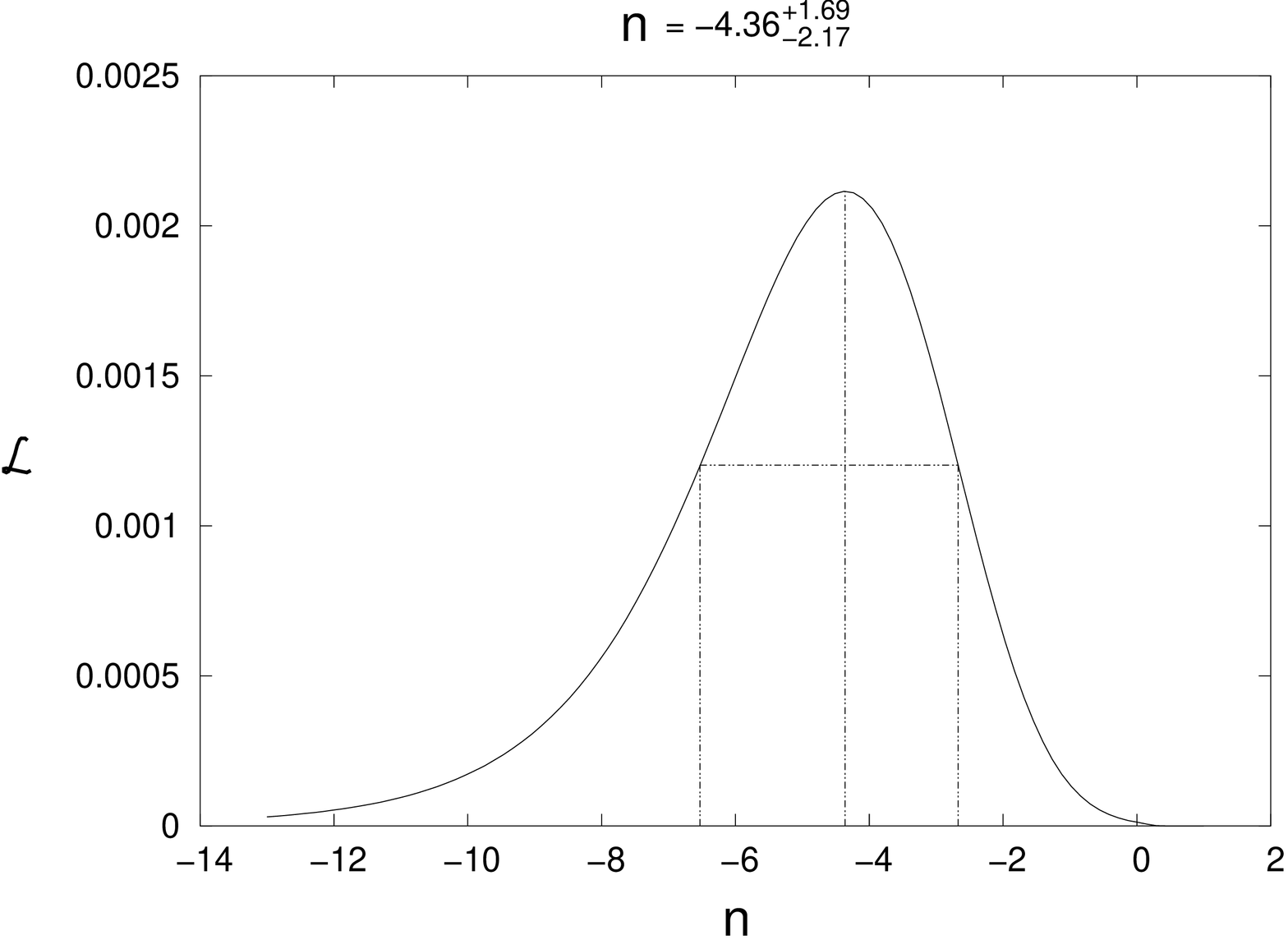} & 
\includegraphics[scale=0.24, angle=270]{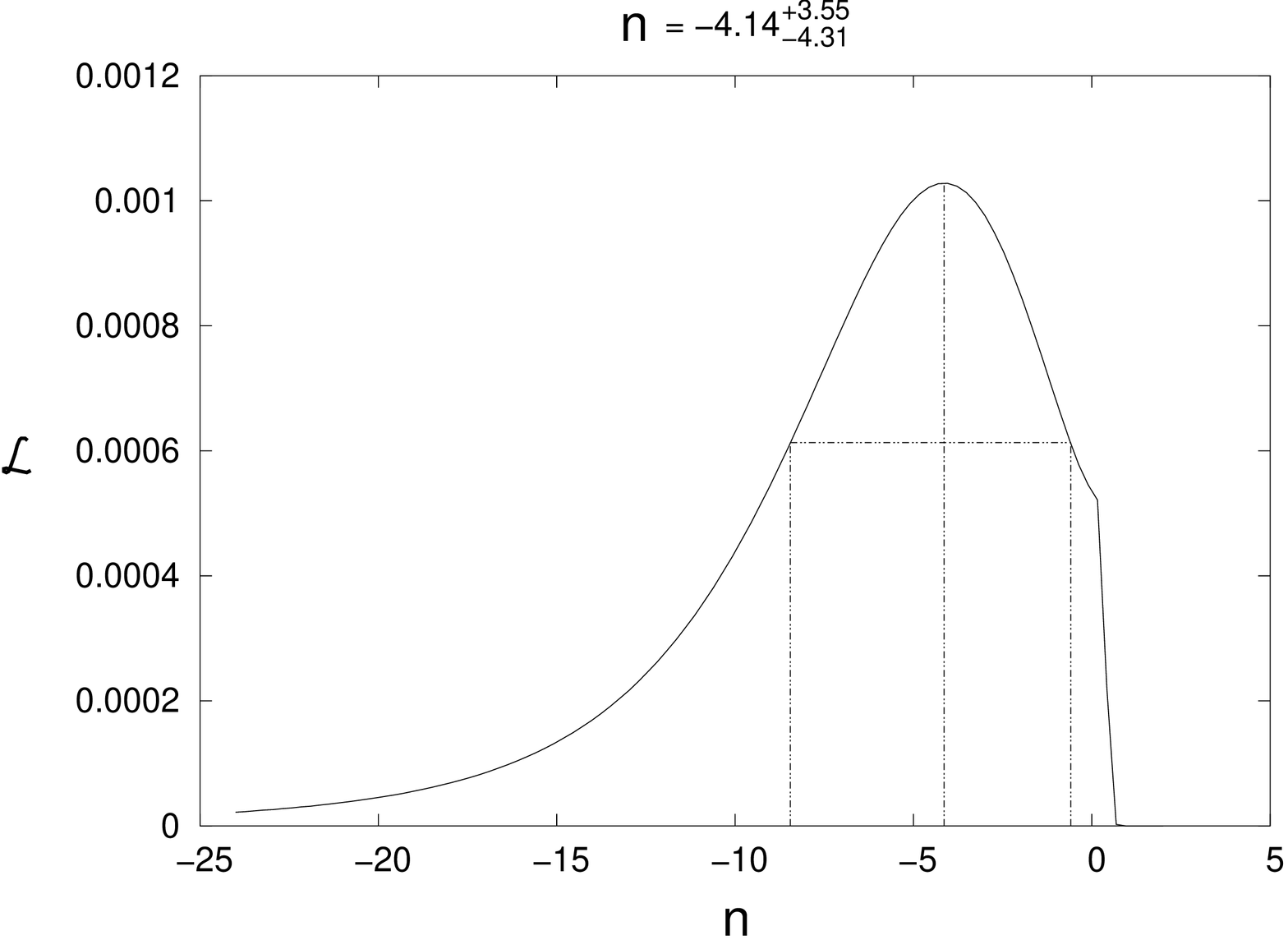} \\ [0.4cm]
%\multicolumn{1}{l}{\mbox{\bf }} & 
\multicolumn{1}{l}{\mbox{\bf (g)}} \\ [-0.5cm]
\includegraphics[scale=0.24, angle=270]{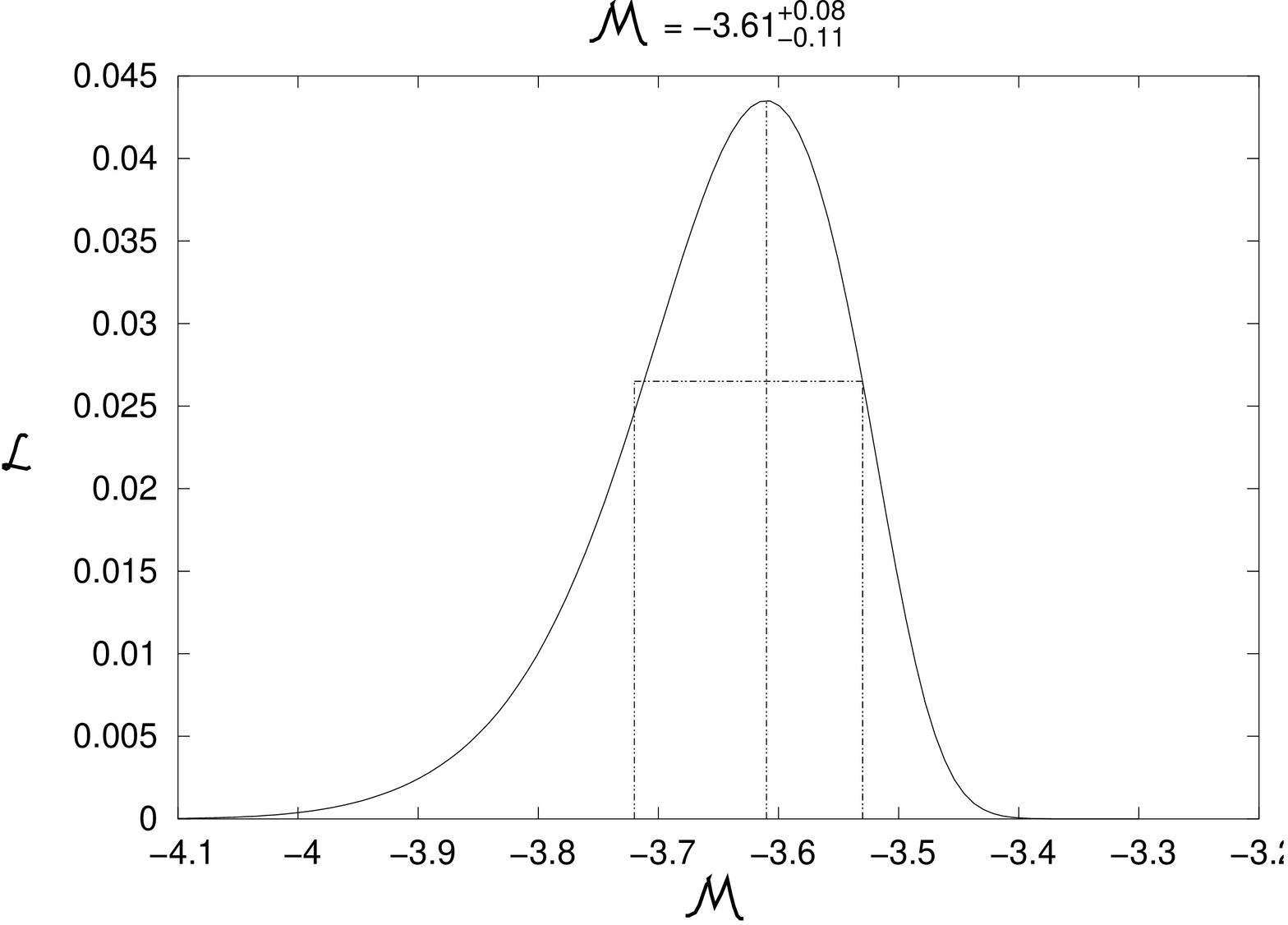}
\end{array}$
\end{center}
\linespread{0.5}
\caption{
The probability density distributions for cosmological parameters $A_{i}$ 
($\Omega_{m,0}$, $\Omega_{{\rm Card},0}$, $n$, $\mathcal{M}$) from the likelihood analysis obtained by the integration of 
the likelihood function $\mathcal{L}(\Omega_{m,0},\Omega_{{\rm Card},0},n, \mathcal{M})$ over remaining 
probability variables $\mathcal{L}(A_{i})=\int_{-\infty}^{\infty}\prod_{k \neq i} dA_{k} \mathcal{L}(A_{j})$.
The cases {\bf (a), (c)} and {\bf (e)} correspond to the analysis with fixed $\mathcal{M}=-3.62$}
\label{stata2}
\end{figure}

Assuming that the measurement uncertainties are Gaussian we can simply calculate the likelihood function
$\mathcal{L}$ from a chi-squared statistics in the following way
\begin{equation}
\mathcal{L}=\frac{1}{N}\exp{\Bigg(-\frac{\chi^{2}}{2}\Bigg)},\quad N=\iiiint_{-\infty}^{\infty}
\exp{\Bigg(-\frac{\chi^{2}}{2}\Bigg)}d\Omega_{m,0}d\Omega_{{\rm Card},0}d\mathcal{M}dn,
\label{erec:6}
\end{equation}
where $N$ is the normalization coefficient given as an integrate of the likelihood function over 
all probability variables.

We use the ``full primary subset'' (Subset 1) of 58 supernovae compiled by Knop et al. \cite{knop03} for the 
statistical analysis. Both the stretch correction and the host-galaxy extinction correction were applied to 
the sample \cite{knop03}. At first the $\chi^{2}$ minimization procedure was prepared for the sample. 
Results of the analysis are presented in the Table~\ref{results1}. Then the three-dimensional space
of estimated parameters ($\Omega_{m,0}$, $\Omega_{{\rm Card},0}$, $n$) was divided into a grid and at each
point of grid the likelihood function was calculated for previously best-fitted value of $\mathcal{M}$ 
parameter. It is important to cover the entire space of estimated parameters where the likelihood function
differs from zero. In our analysis the exploded space included ranges of parameters: $\Omega_{m,0}\in[0:1]$,
$\Omega_{{\rm Card},0}\in[0:1]$, $n\in[-13:1]$. 

The confidence levels for pairs $(\Omega_{m,0},n)$, obtained
through the integration of the likelihood function $\mathcal{L}(\Omega_{m,0},\Omega_{{\rm Card},0},n)$
over the $\Omega_{{\rm Card},0}$ variable, are drawn in Fig.~\ref{stata1}{\bf (a)}.

The probability density distributions for cosmological parameters $A_{i}$ 
($\Omega_{m,0}$, $\Omega_{{\rm Card},0}$, $n$) from the likelihood analysis, obtained by the integration of 
the likelihood function $\mathcal{L}(\Omega_{m,0},\Omega_{{\rm Card},0},n)$ over remaining probability 
variables $\mathcal{L}(A_{i})=\int_{-\infty}^{\infty}\prod_{k \neq i} dA_{k} \mathcal{L}(A_{j})$, 
are presented in Fig.~\ref{stata2}{\bf (a), (c)} and {\bf (e)}.

\begin{figure}[!ht]
\begin{center}
$\begin{array}{c@{\hspace{0.2in}}c@{\hspace{0.2in}}c}
\multicolumn{1}{l}{\mbox{\bf (a)}} & 
\multicolumn{1}{l}{\mbox{\bf (b)}} &
\multicolumn{1}{l}{\mbox{\bf (c)}} \\ [-0.0cm]
\includegraphics[scale=0.45, angle=0]{62.eps} & 
\includegraphics[scale=0.45, angle=0]{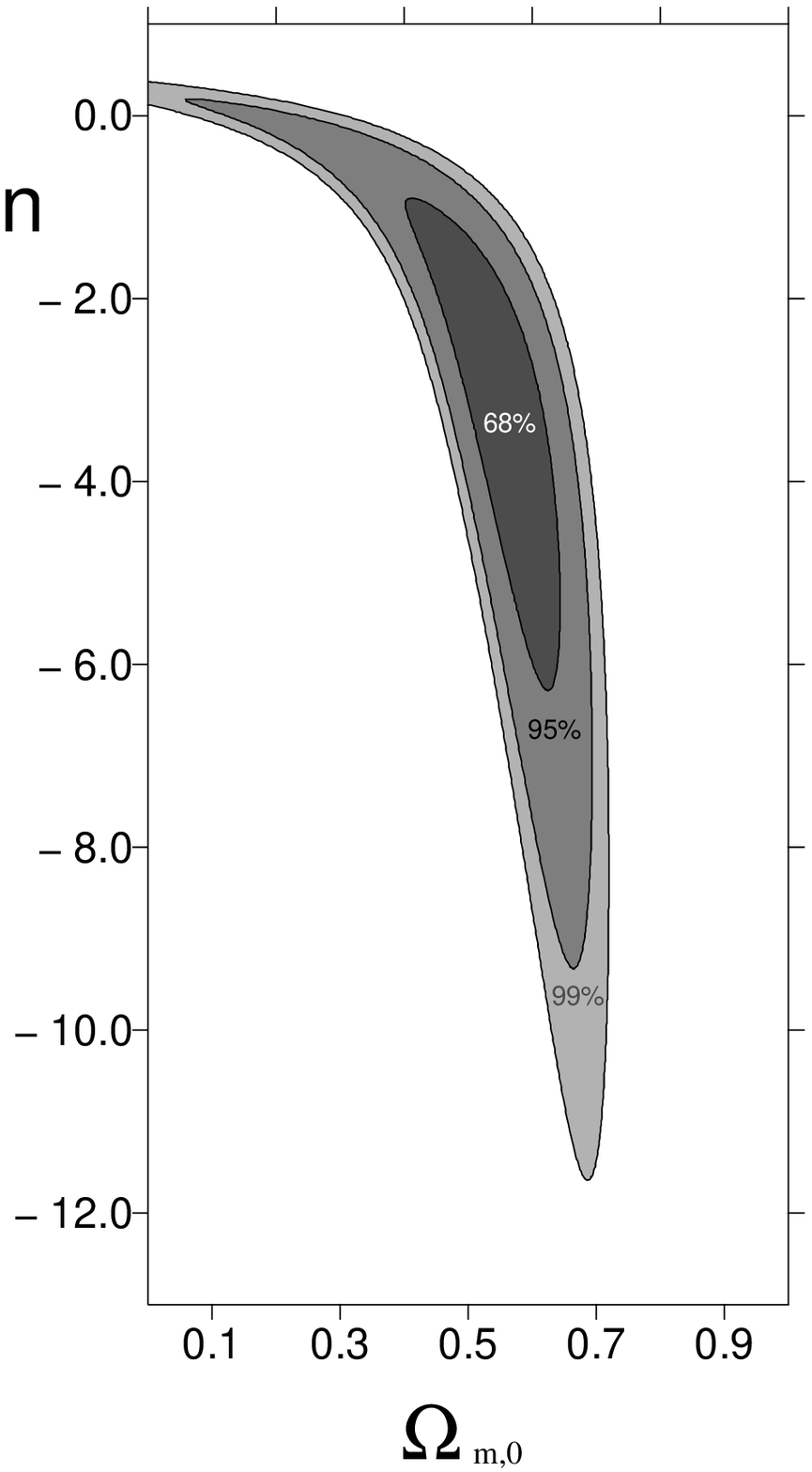} & 
\includegraphics[scale=0.45, angle=0]{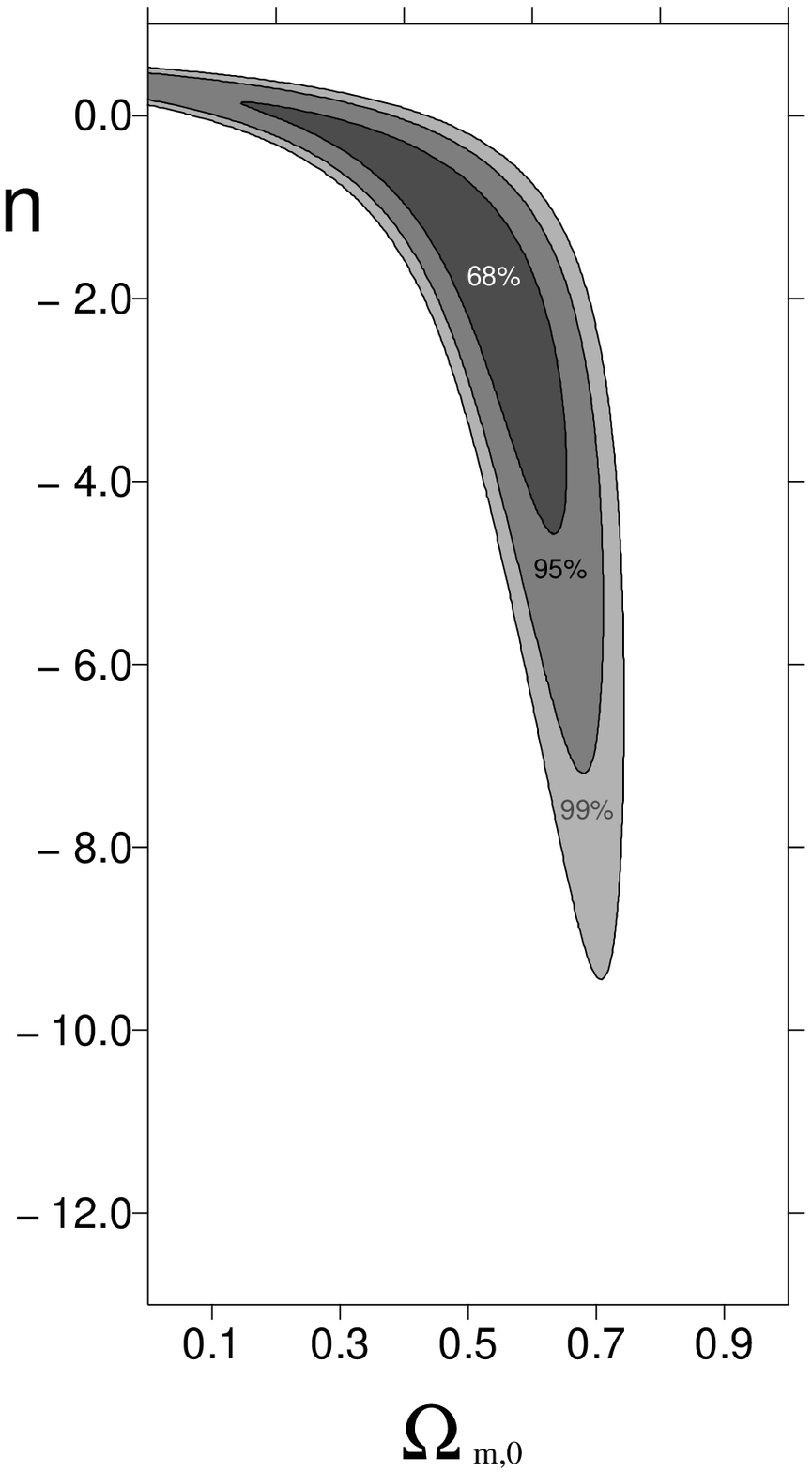}
\end{array}$
\end{center}
\linespread{0.5}
\caption{Confidence levels on the plane $(\Omega_{m,0},n)$ for the flat Cardassian model
and different values of $\mathcal{M}$ parameters {\bf (a)} $\mathcal{M}=3.62$ (best fitted value for the 
model), {\bf (b)} $\mathcal{M}=3.58$ and {\bf (c)} $\mathcal{M}=3.54$ (best fitted value for curved model 
with non-relativistic matter and cosmological constant). Note that results of the statistical analysis
strongly depend on the value of $\mathcal{M}$ parameter.}
\label{stata4}
\end{figure}

As it is shown in Fig.~\ref{stata4} the likelihood function strongly depend on the value of $\mathcal{M}$
parameter so it is very important to choose it as a best-fitted value for the model or (for a more robust 
determination) to treat $\mathcal{M}$ as an additional probability variable and integrate full likelihood 
function also over $\mathcal{M}$. The analysis with marginalization over $\mathcal{M}$ was prepared in 
the similar way. The results of the analysis were presented in the Table \ref{results1}. The confidence 
levels on the plane $(\Omega_{m,0},n)$ and one-dimensional probability density distributions for 
the cosmological parameters $\Omega_{m,0}$, $\Omega_{{\rm Card},0}$, $n$, $\mathcal{M}$ are shown in 
Fig.~\ref{stata1}{\bf (b)} and Fig.~\ref{stata2}{\bf (b),(d),(f),(g)} respectively.
Of course, in this case, the uncertainties of the cosmological parameters are bigger because of one 
additional estimated parameter. 

\begin{figure}[!ht]
\begin{center}
$\begin{array}{c@{\hspace{0.2in}}c}
\multicolumn{1}{l}{\mbox{\bf (a)}} & 
\multicolumn{1}{l}{\mbox{\bf (b)}} \\ [0.2cm]
\includegraphics[scale=0.3, angle=270]{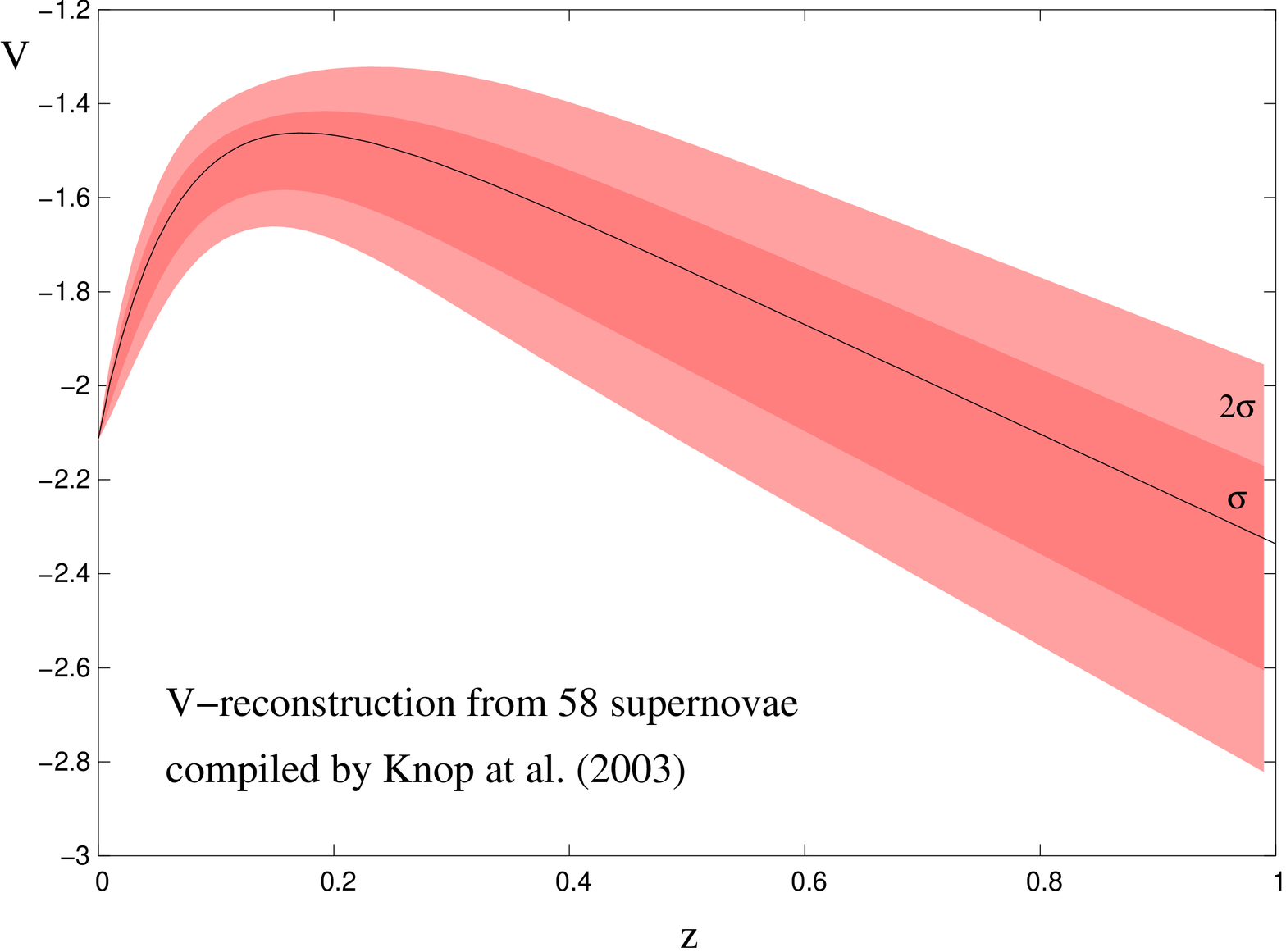} & 
\includegraphics[scale=0.3, angle=270]{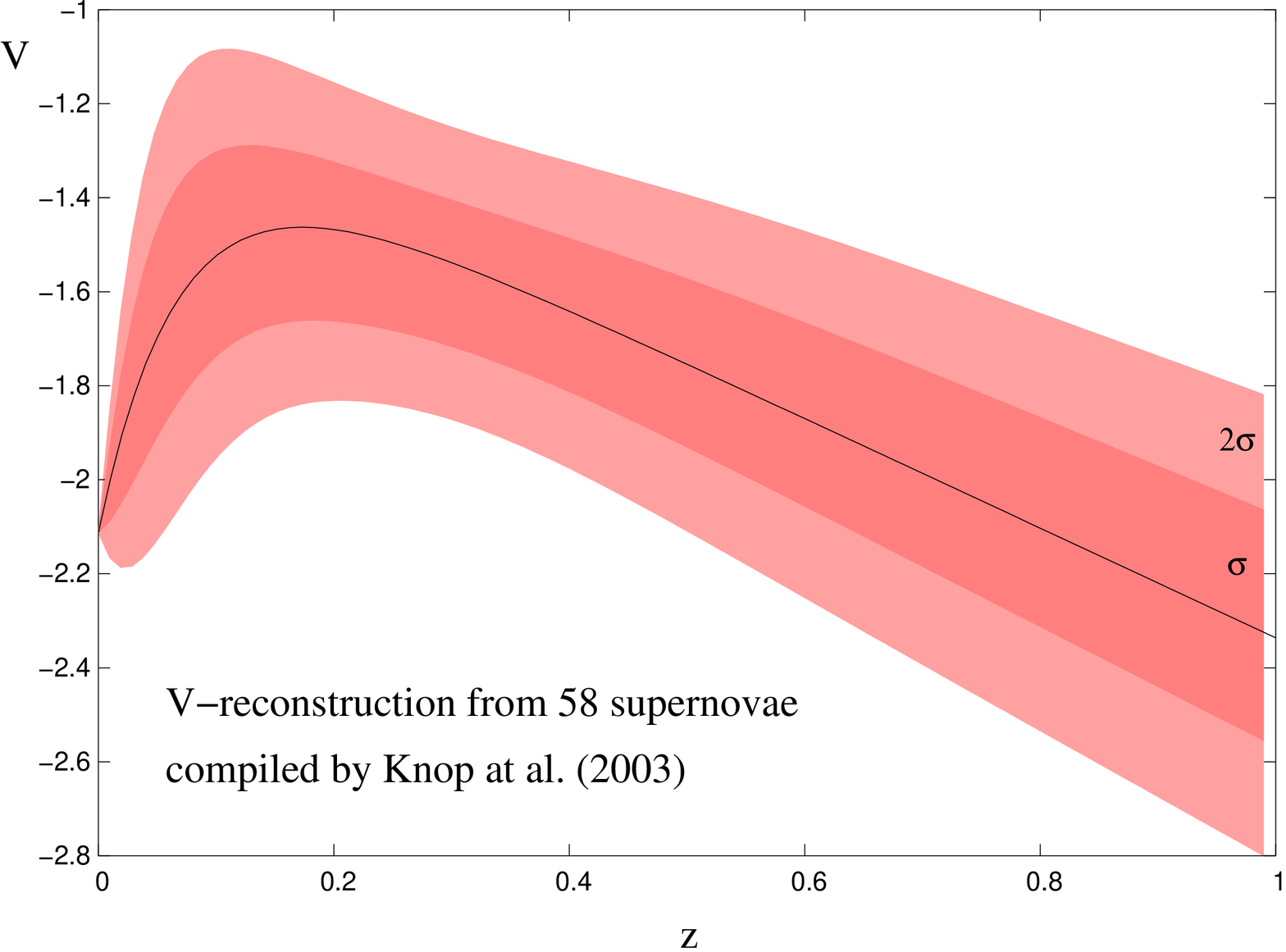}
\end{array}$
\end{center}
\linespread{0.5}
\caption{Confidence levels ($1\sigma$ and $2\sigma$) for the potential function $V(z)$ given by the formula
(\ref{erec:4}) obtained from the likelihood method (see Table~\ref{results1}). The case {\bf (a)} corresponds
to the analysis with fixed $\mathcal{M}=-3.62$. The best fit is represented by the solid black line.}
\label{stata3}
\end{figure}

It is well known that there is a simple interpretation of the Cardassian term as an additional 
noninteracting fluid with pressure $p=[n(1+w)-1]\rho$, where the standard term is obtained from the fluid 
$p=w \rho$ and $w=\mathrm{const}$. In the special case of dust we obtain $p=(n-1)\rho$. Let us note
(see Fig.~\ref{stata3}{\bf (a)}) that the supernovae data favor the high density matter model of the 
universe with $n<0$ at the $99\%$ confidence level (it is not truth for the cases {\bf (b)} and {\bf (c)}). 
From the previously presented interpretation of the Cardassian model as a FRW model filled with a mixture 
of noninteracting fluids we can conclude that the universe is now dominated by the mater with a 
super-negative pressure $p<-\rho$. At first the fact was noticed by T.~Roy Choudhury and T. Padmanabhan 
\cite{choudhury03}.

\begin{figure}[!ht]
\begin{center}
\includegraphics[scale=0.35,angle=270]{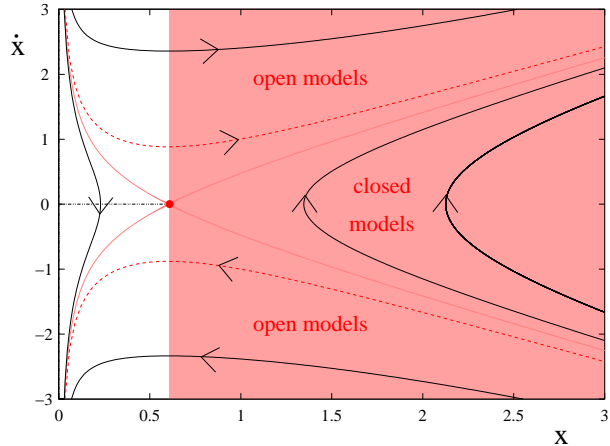} 
\end{center}
\linespread{0.4}
\caption{Phase space for the system (\ref{eq:8}), where the potential function $V[x(z)]$ given by the 
equation (\ref{erec:4}) has been reconstructed from the SNIa observational data \cite{knop03}.
The shaded domain of phase space is the domain of accelerated expansion of the universe. The dashed line 
represents the flat model trajectory which separates the regions with negative and positive curvature.} 
\label{stata5}
\end{figure}

We learn from the previous dynamical systems analysis that all what is required for unique determination
of the Cardassian models contain the diagram of the potential function. Since the Hubble function
is related to the luminosity distance (for the flat model) by the simple relation
\begin{equation}
H(z)=\Bigg[\frac{d}{dz}\Bigg(\frac{d_{L}(z)}{1+z}\Bigg)\Bigg]^{-1}
\label{erec:7}
\end{equation}
it is possible to determine both the potential function for the system (or equivalent $\rho_{\rm eff}(z)$)
and then quintessence parameter $w(z)=p_{\rm eff}/\rho_{\rm eff}$. In our previous papers 
\cite{szydlowski03,szydlowski03c} it was demonstrated that the reconstruction of the potential function
can be performed for the general class of FRW models with an equation of state in the form $p=w(a)\rho$.
Such an approach to the reconstruction of the dynamics, in terms of the potential (or energy density), directly
from observations in the model independent manner has two advantages:

\noindent
1) the dynamics can be derived only from the single function and the description has a natural 
interpretation -- particle-like description;

\noindent
2) the reconstruction of $V[a(z)]$ is a simpler procedure then the probe of the dark energy in term of
quintessence parameter $w(z)$, for reasons pointed out by Maor, Brunstein and Steinhardt \cite{maor01}.

Because of many problems in determining the quintessence parameter $w(z)$ we propose the reconstruction of
$V[a(z)]$ \cite{szydlowski03,szydlowski03c} as a simpler alternative to description the dynamics of 
quintessential cosmology. Later some authors \cite{wang04}, for the same reasons, use a rescalled quantity
of the potential, i.e., $\rho(z)=-6V(z)(1+z)^{2}$ having the sense of effective energy density.
Advantage of use of the potential function is that the dynamics can be completely determined. 
Our paper shows the advantage of using $V(z)$ instead of $w(z)$ in investigation of the dynamics of
quintessential models.

Note that both deceleration parameter $q$ and quintessential parameter $w$ are simply related to the
elasticity of the potential with respect to the scale factor $I_{V}(a)$.
\begin{gather}
q=-\frac{1}{2}I_{V}(a),\quad w(t)=-\frac{1}{3}(I_{V}(a)+1), \nonumber \\
\dot{w}=-\frac{1}{3}\dot{I_{V}}(a),
\end{gather}
where $I_{V}=\partial \ln{V}/ \partial \ln{a}$ is the elasticity parameter.

Because trajectories of the system lie on the zero energy level (Hamiltonian constraint) we obtain 
\begin{equation}
V(z)=-\frac{1}{2}(1+z)^{-2}\Bigg[\frac{d}{dz}\Bigg(\frac{d_{L}(z)}{1+z} \Bigg)\Bigg]^{-2},
\label{erec:8}
\end{equation}
where $d_{L}(z)$ is fitted by using the Cardassian model itself rather then the polynomial ansatz
(as in our previous work) because it gives better results. 

Therefore from the above formula we can simply obtain the effective energy density
\begin{equation}
\rho_{\rm eff}(z)=3\Bigg[\frac{d}{dz}\Bigg(\frac{d_{L}(z)}{1+z} \Bigg)\Bigg]^{-2}.
\label{erec:9}
\end{equation}

Note that in  both last formulas only first derivatives of $d_{L}(z)$ appear while $w(z)$ depends on the
$d_{L}(z)$ through the second order differentiation.

In the Fig.~\ref{stata5} we present the best fit for the reconstructed potential function of the Cardassian 
scenario and the confidence levels ($1\sigma$ and $2\sigma$) for the potential obtained from the likelihood 
statistical analysis while the phase portrait for the reconstructed potential function is shown in 
Fig.~\ref{stata4}. In the statistical analysis we consider the polynomial function and the power low
function. The lowest $\chi^{2}$ was obtained for the latter so we assume the power low ansatz for
the reconstruction of the potential function.

It is very interesting that from the reconstructed potential we can immediately obtain the phase portrait.
It is a consequence of the fact that the potential uniquely determines the dynamics of the Hamiltonian flow.
This portrait is representing the evolutional path of the Cardassian model for all initial conditions.
Note that the phase portraits from Fig.~\ref{fig5}{\bf (a),(b),(c)} are topologically equivalent to that 
obtained from the reconstruction (see Fig.~\ref{stata5}).

\section{On some other special features of Cardassian models}

It seems to be reasonable to require the phase portrait equivalence to be fulfilled. It means that any
model of the universe should describe its dynamics without contradiction with astrophysical data.
Therefore, the acceptable model should have phase portrait (qualitative dynamics) equivalent to the
phase portrait reconstructed from observations.

In this section we would like to present how Cardassian models can explain some cosmological problems
\cite{barrow03}. The Cardassian models are able to provide a natural explanation for our observations
of the universe which is almost isotropic and flat. In standard FRW cosmology with perfect fluid we can 
observe that at large scale factor $a$ the curvature term falls off faster than $a^{-3(1+w)}$ term whenever
$w<-1/3$, i.e., some special kind of fluid is required for explanation of the flatness problem.
By contrast, in the Cardassian scenario the flatness problem is naturally solved because the $ka^{-2}$ term
falls off faster than the $Ba^{-3(1+w)n}$ term provided $n<2/(3(1+w))$. Therefore the period of 
dust-dominated evolution of the universe with $n<2/3$, which satisfies following Cardassian assumption,
can solve the classical flatness problem.

Let us consider the class of anisotropic but homogeneous cosmological models, namely Bianchi model with
perfect fluid, at late time of evolution. In these models anisotropy rate arises from scalar anisotropy,
which falls off no slower than $\sigma^{2} \propto a^{-2}$ at late time, where $a$ is an average scale factor.
In the homogeneous, but anisotropic flat Cardassian cosmological model the basic equation is modified by 
addition of a $\sigma^{2}$ term to its right-hand side. Therefore, the isotropy problem is solved by 
accelerating universe with $0 \leq w \leq 2/3$ because scalar anisotropy $\sigma^{2}$ falls off faster
than the $a^{-3(1+w)}$ term. If we consider Cardassian model with $n<2/(3(1+w))$ then the anisotropy term
can not dominate the expansion. Thus the anisotropy problem can be solved in a dust-dominated universe by 
an interval of Cardassian evolution if $n<2/3$.
Of course, in the standard cosmology if $\rho+3p<0$ (strong energy condition is violated) the matter term
falls off more slowly than $\sigma^{2}$ as $t \rightarrow \infty$, i.e., anisotropy is damping during the 
accelerating phase of the evolution whereas the Cardassian scenario offers a natural explanation of 
anisotropy problem without assumption of a special form of dark matter. It is a consequence that Cardassian 
models solve the flatness problem. Therefore these models do not offer a new solution of the isotropy
problem. This situation in some sense is analogical to that which take place in the context of VSL 
cosmologies \cite{szydlowski03b}. However let us note that VSL models in contrast to the Cardassian models
can solve the horizon problem in the past.

Let us now consider some general situation in which we have $w=w[a(t)]$. Such a case appear if we consider
both matter and radiation for example. Then the following theorem establish the sufficiently condition for 
the presence of particle horizon. 

When all events whose coordinates in the past are located beyond some distance $d_{H}$ then can never 
communicate with the observer at the coordinate $r = 0$ (in R-W metric). We can define the distance
$d_{H}$ as the past event horizon distance
\begin{equation}
d_{H}(t) = a(t) \int_{t_{0}}^{t}\frac{dt'}{a(t')}c = a(t)I.
\end{equation}
Of course, whenever $I$ diverges as $t_{0} \rightarrow 0$, there is no past event horizon in the spacetime 
geometry. Then it is in principle possible to receive signals from sufficiently early universe from any 
comoving particle like a typical galaxy. If the $t'$ integral converges for $t_{0} \rightarrow 0$ then our 
communication with observer at $r=0$ is limited by what Rindler has called a particle horizon.
The particle horizon will be present if energy density is growing faster then $a^{-2-\epsilon}$ as 
$a \rightarrow 0$ ($\epsilon \geq 0$). Therefore if $\dot{a}>a^{-\epsilon/2}$ ($\dot{a}>A$) then there is 
a particle horizon in the past. 

Our investigation of the particle horizon here is independent
of any specific assumption about the behavior of $a(t)$ near the singularity or of specific form of the 
equation of state. If one assumes a linear equation of state $p = w \rho$ and $w = {\rm const}.$, then 
Friedman's equations imply the following behavior for $a(t) \backsimeq (t)^{\frac{2}{3(\gamma+1)}}$ near 
the singularity $t = 0$. The integral $I$ would thus diverge only if $\gamma < -1/3$, i.e., only if the 
pressure $p$ becomes negative. This is the condition for solving the horizon problem and it is identical to 
that for the solution of the flatness problem ($\rho/3$ term will dominate the curvature term in a 
long time evolution). We can also show here that the integral $\int \frac{dt}{a(t)}$ would diverge only if 
the pressure of the cosmological fluid takes negative values in the general case of $w(a(t))$. 
Conservation condition can be rewritten in the form
\begin{equation}
a^{3}dp = d[\rho a^{3}(1+w(a))].
\end{equation}
We can verify that boundedness of $\dot{a}(t)$ means also that $\rho a^{2}$ remains bounded near the
singularity. Therefore, $\rho a^{3} \rightarrow 0$ as $t \rightarrow 0$. By integrating both
sides of the equation written above from $0$ to $t$ we obtain 
$-3\int_{0}^{t}w(a)\rho a^{2}\dot{a}dt = a^{3}\rho(t) \geqslant 0$. \\
Consequently, $w(a)$ must assume a negative value without any specific assumption about the equation of 
state.

\section{Conclusions}

In this paper we apply the qualitative cosmology analysis to the general class of Cardassian models which 
have become rather an alternative to the dark energy models. We show that while the application of 
qualitative methods allows to reveal some unexpected properties like the big-rip singularities, the SNIa
data uniquely reconstruct the phase plane in the model independent way. The reconstruction of this phase 
plane should be treated as a necessary condition for any model which wants to explain the present acceleration
of the universe.

Theory of dynamical systems which offer a possibility of investigating the space of all solutions for all
admissible initial conditions is used in analysis of Cardassian models. We demonstrate a simple method of
reducing of the dynamics to certain two-dimensional dynamical system. One of the features of this reduction
is the possibility of representing the model as a Hamiltonian system in which the properties of the 
potential function can serve as a tool for qualitative classification of all possible evolution scenarios.
It is shown that some important features like resolution of the flatness problem or horizon problem and
its degree of generality can be visualized as domains on the phase plane. Then one is able to see how large
is the class of solutions (labelled by the initial conditions) leading to the desired property, 
i.e., this class is generic or non-typical.

Applying the techniques of the particle-like description, developed by us earlier, to a new data set
from the Knop's sample we show that the Cardassian model with $n<0$ is favored by the data on the 
$99\%$ confidence level. On the other hand it is just the case of appearance of unexpected big-rip
singularities in the future evolution of the model.

Although the developed formalism is mainly adopted to the investigation of the Cardassian models it can be 
useful to include additionally dissipative effects of bulk viscosity to the model. Moreover for these class
of models de Sitter solution is admissible as a global attractor in the future.

The main results are the following:

\begin{enumerate}
\item
We show the effectiveness of using of dynamical system methods in analysis of the wide class of Cardassian 
models and their generalizations.
\item
The unwanted big-rip singularities at a finite time were detected as a generic for the case $n<0$ 
(a global attractor in the future).
\item
The exact solutions and some duality relations were found.
\item
Additionally we showed that the existence of this kind of singularities is acceptable on the $99\%$ confidence 
level from Knop's sample of SNIa data.
\item
The particle-like approach was adopted to describe the dynamics as well as the potential function was
reconstructed from the observations. In the fitting procedure the power law function are used because
it gives the lowest value of $\chi^{2}$.
\item
Because of interpretation of the Cardassian term as an effect of additional fluid one can claim that if FRW
equation is ``correct'' then phantom fields are required at the $99\%$ confidence level.
\end{enumerate}

Finally, our study showed that the Cardassian model with $n<0$ is strongly favored by the Knop's 
SNIa data (without any priors on matter content). But, this model predicts rather unexpected 
future of the universe -- a big-rip singularity. However, the estimation of characteristic time 
of the big-rip singularity will be a subject of our next paper.

\begin{acknowledgments}
The paper was supported by KBN grant No. 1P03D 003 26. Authors are very gratefull to Mariusz Dabrowski 
and Adam Krawiec for discussion and comments.
\end{acknowledgments}

\end{document}